\documentclass[a4paper,11pt]{article}

\usepackage{jheppub}

\usepackage[T1]{fontenc} 
\usepackage{dsfont}
\usepackage{framed}

\usepackage{caption}
\usepackage{subcaption}
\usepackage{dsfont}
\usepackage{tikz}
\usepackage{array}
\usepackage{makecell}
\usepackage[aligntableaux=center,boxsize=1.3em]{ytableau}

\allowdisplaybreaks[1]

\def\be#1\ee{\begin{align}#1\end{align}}

\def\({\left(}
\def\){\right)}
\def\[{\left[}
\def\]{\right]}
\def\<{\langle}
\def\>{\rangle}

\def\nn{\nonumber\\}

\def\rm{\mathrm}
\def\bb{\mathbb}
\def\cal{\mathcal}

\def\pa{\partial}
\def\ra{\rightarrow}

\def\xn{x_{\perp}}

\def\yn{y_{\perp}}

\def\a{\alpha}
\def\b{\beta}
\def\g{\gamma}
\def\G{\Gamma}
\def\d{\delta}
\def\D{\Delta}
\def\e{\epsilon}

\def\i{\iota}

\def\l{\lambda}

\def\m{\mu}
\def\n{\nu}
\def\x{\xi}

\def\p{\pi}

\def\f{\phi}
\def\F{\Phi}

\def\s{\sigma}

\def\Ps{\Psi}

\def\h{\hat}

\begin{document}

\title{Boundary anomalous dimensions from BCFT: \\
\boldmath{$\phi^{3}$} theories with a boundary 
and higher-derivative generalizations}

\author[a,b]{Yongwei Guo}
\author[a]{and Wenliang Li}
\emailAdd{liwliang3@mail.sysu.edu.cn}

\affiliation[a]{School of Physics, Sun Yat-sen University, Guangzhou 510275, China}
\affiliation[b]{Shing-Tung Yau Center and School of Physics, Southeast University, Nanjing, 210096, China}

\abstract{We consider the bulk $\f^3$ deformation of the free boundary conformal field theory in the $\e$ expansion.
We determine the leading corrections to the scaling dimensions of boundary fundamental operators and some boundary operator expansion coefficients.
Our procedure combines the conformal multiplet recombination with the boundary crossing symmetry.
The results cover both the single field case and the multi-field case with $S_{N+1}$ global symmetry,
which are associated with the Yang-Lee model and the $(N+1)$-state Potts model respectively.
These semi-infinite models describe branched polymers, percolation, and spanning forest at a surface.
We generalize these results to some higher derivative theories.
In addition, we study the $\f^{2n+1}$ theories with $n>1$, but only obtain some boundary operator expansion coefficients.}
	
\maketitle 

\section{Introduction}

The conformal field theory (CFT) with a $\f^3$ interaction has many interesting physical applications.
For the single field case, the $\f^3$ theory describes the critical behaviors of Yang-Lee edge singularity, 
as well as branched polymers 
\footnote{Branched polymers consist of linear polymers connected by branch points. 
A linear polymer is a chain of monomers with vertices repulsing each other.
In the many-monomer limit, branched polymers exhibit critical behavior which belongs to the same universality class as the random field $\f^3$ model.
This model has emergent Parisi-Sourlas supersymmetry and corresponds to the pure $\f^3$ theory in two dimensions lower.} 
in two dimensions higher due to the Parisi-Sourlas dimensional reduction \cite{Parisi:1980ia,brydges2003branched,cardy2003lecture,Kaviraj:2021qii,Kaviraj:2022bvd}.\footnote{See \cite{Rychkov:2023rgq} for a review on the more recent developments. The Parisi-Sourlas mechanism \cite{Parisi:1979ka,Kaviraj:2019tbg} may break down for the $\f^4$ case \cite{Kaviraj:2021qii,Fytas:2017fvr,Fytas:2019zdk,Fytas:2019byz,Kaviraj:2020pwv,Fytas:2023izo}.}
For the multi-field case with $S_{N+1}$ symmetry, the pure $\f^3$ theory is related to the phase transition of the $(N+1)$-state Potts model.
Here $S_{N+1}$ is the symmetric group, i.e., the group of all permutations of $N+1$ elements.
Two particularly interesting limits of the Potts model are $N\ra 0$ and $N\ra -1$, which correspond to the percolation problem and spanning forest \cite{Jacobsen:2003qp,Caracciolo:2004hz,Deng:2006ur}.
In this work, we consider the boundary conformal field theory (BCFT) associated with the aforementioned physical systems in the presence of a boundary \cite{PhysRevB.19.6295,carton1980surface,KDeBell1980,AChristou1986,Janssen1995}.
See also \cite{Diehl:1996kd} for a review on boundary critical phenomena. 
Consider the BCFT on the Euclidean upper half space
\be
\bb R^d_+=\{(x_\parallel^\m,x_\perp):x_\parallel^\m\in\bb R^{d-1},\, x_\perp\geq 0\}
\,.
\ee
Together with a Dirichlet or Neumann boundary condition, the free BCFT at the upper critical dimension $d_{\text u}=6$ describes the so-called ordinary or special transition.\footnote{This identification may be more subtle in the interacting theory if the boundary condition is scale dependent \cite{Diehl:2020rfx}.}
The surface critical exponent $\eta_{\parallel}$ are determined by the scaling dimension of the boundary fundamental operator $\F$.\footnote{The relation is usually given by $\eta_\parallel=2\D_\F-d+2$. See however \cite{Janssen1995} for an exception in the adsorption-transition of branched polymers at a boundary.}
\footnote{There is another independent surface critical exponent at the special transition, namely the crossover exponent.
This exponent is related to the scaling dimension of the boundary composite operator $\F^2$, which we will not discuss in the present paper.}
Below the upper critical dimension $d_{\text u}$, there are corrections to the boundary scaling dimension $\hat \D_\F$, 
generated by the $\f^3$ interaction.\footnote{The lowest dimension of an operator associated with a boundary interaction is $6+O(\e)$.
This is higher than the dimension of the boundary, which is $5+O(\e)$.
So there is no mixing between the bulk and boundary interactions in the $\f^3$ theory.
This conclusion is also true for higher derivative $\f^3$ theories.}
Furthermore, we will extend the discussion to theories with $\mathbb Z_2$ odd interactions $\phi^{2n+1}$, where $n>1$ is an integer.
\footnote{See \cite{Guo:2025edk} for the case of the $\f^{2n}$ deformation.}
In what follows, we briefly explain the main methods used in this work.

In recent years, the multiplet recombination method \cite{Rychkov:2015naa} has been used to study various conformal fixed points in the $\e$ expansion \cite{Rychkov:2015naa,Basu:2015gpa,Yamaguchi:2016pbj,Ghosh:2015opa,Raju:2015fza,Gliozzi:2016ysv,Roumpedakis:2016qcg,Soderberg:2017oaa,Behan:2017emf,Gliozzi:2017hni,Gliozzi:2017gzh,Nishioka:2022odm,Nishioka:2022qmj,Antunes:2022vtb,Guo:2023qtt,Guo:2024bll,Antunes:2024mfb,DeCesare:2025ukl,DeCesare:2026dwm} (see also \cite{Safari:2017irw,Nii:2016lpa,Hasegawa:2016piv,Hasegawa:2018yqg,Skvortsov:2015pea,Giombi:2016hkj,Giombi:2017rhm,Giombi:2016zwa,Codello:2017qek,Codello:2018nbe,Antipin:2019vdg,Vacca:2019rsh,Giombi:2020rmc,Giombi:2020xah,Dey:2020jlc,Giombi:2021cnr,Safari:2021ocb,Zhou:2022pah,Bissi:2022bgu,Herzog:2022jlx,Giombi:2022vnz,SoderbergRousu:2023pbe,Herzog:2024zxm}).
For a $\f^3$ theory, the free multiplets of bulk primaries $\f_{\text f}$ and $\f^2_{\text f}$ recombines into a long multiplet in the interacting theory, which can be represented schematically as
\be
\{\f\}_{\text{interacting}}\approx\{\f\}_{\text {free}}+\{\f^2\}_{\text {free}}
\,.
\ee
We use the subscript f to indicate free-theory expressions throughout this paper.
In the limit $\e\rightarrow 0$, the deformed version of the descendant $\Box\f$ is proportional to $\f^2_{\text f}$,
\be
\lim_{\e\ra 0}\e^{-1/2}\Box\f\propto\f^2_{\text f}
\,,
\ee
where the exponent $-1/2$ is determined in by the bulk analysis (see e.g. \cite{Guo:2024bll}).
The multiplet recombination method bypasses the diagrammatic calculations and provides an axiomatic approach to determine the conformal data.

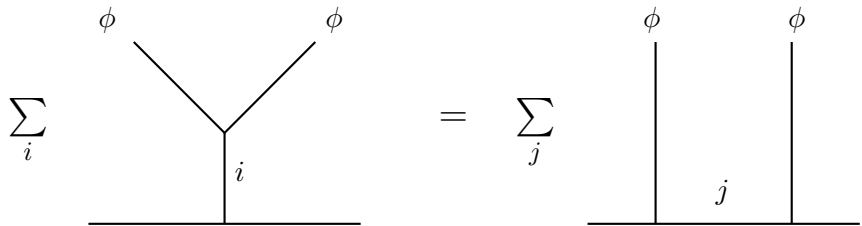
\begin{figure}
\center
\begin{tikzpicture}[scale = 1.2]
\draw[black, thick] (-4,2) -- (-3,1);
\draw[black, thick] (-2,2) -- (-3,1);
\draw[black, thick] (-3,1) -- (-3,0);
\draw[black, thick] (-4.5,0) -- (-1.5,0);

\draw[black, thick] (1.75,2) -- (1.75,0);
\draw[black, thick] (3.25,2) -- (3.25,0);
\draw[black, thick] (1,0) -- (4,0);

\draw (-5.5,1.5) node [anchor=north west]   {\Large $\sum\limits_i$};
\draw (-0.75,1.5) node [anchor=north west]   {\Large $=\hspace{1em}\sum\limits_j$};

\draw (-4.5,2.5) node [anchor=north west]   {$\phi$};
\draw (-2,2.5) node [anchor=north west]   {$\phi$};

\draw (1.5,2.5) node [anchor=north west]   {$\phi$};
\draw (3.1,2.5) node [anchor=north west]   {$\phi$};

\draw (-3,0.8) node [anchor=north west]   {$i$};
\draw (2.3,0.6) node [anchor=north west]   {$j$};

\end{tikzpicture}
\caption{The crossing equation for the bulk two-point function $\<\phi(x)\phi(y)\>$.
The left hand side (LHS) is the bulk channel, where we consider the bulk OPE of $\phi(x)$ and $\phi(y)$.
The correlator $\<\phi(x)\phi(y)\>$ is expanded in terms of a sum of bulk one-point functions.
On the right hand side (RHS), each $\f$ is decomposed into a sum of boundary operators, 
so $\<\phi(x)\phi(y)\>$ becomes as a sum of boundary two-point functions.}
\label{crossing equation}
\end{figure}

The multiplet recombination is different from the conformal bootstrap methods, in which crossing
symmetry leads to powerful constraints on conformal field theories (CFTs). 
As suggested in \cite{Rychkov:2015naa}, it should be interesting to combine these two different methods to further constrain the CFT data. 
This idea can be simplified in the context of boundary conformal field theory (BCFT),\footnote{See \cite{Andrei:2018die} for a recent review on BCFT.}
as the boundary crossing equation for a two-point correlator contains only a single cross ratio \cite{Liendo:2012hy}.
\footnote{See \cite{Giombi:2020xah,Herzog:2022jlx} also for the cases of real projective CFT and fermionic BCFT, which make use of both the equations of motion and crossing symmetry. }
As illustrated in Fig. \ref{crossing equation}, we can expand the bulk two-point correlator $\<\f(x)\f(y)\>$ in two ways.
One is associated with the bulk operator product expansion (OPE), while the other is related to the boundary operator expansion (BOE).
The two different expansions should give the same result for $\<\phi(x)\phi(y)\>$, leading to the boundary crossing equation.

Our goal is to determine the anomalous dimension $\hat\g_\F$ of the boundary fundamental operator $\F$.
Let us explain the problems of using only the multiplet recombination method or boundary crossing symmetry.
A naive application of the multiplet recombination does not fully determine $\hat\g_\F$.
More specifically, according to the previous work on the $\f^{2n}$ case \cite{Guo:2025edk}, 
one may try to study the bulk-boundary two-point function $\<\f(x)\F(\h y)\>$.
(The hat indicates coordinates on the boundary.)
This does not work because the multiplet recombination relates it to the vanishing free correlator $\<\f^2(x)\F(\h y)\>_{\text f}$.
In other words, the multiplet recombination only leads to a constraint on the order of the anomalous dimension.
On the other hand, if one considers the crossing symmetry for $\<\f(x)\f(y)\>$, 
the boundary crossing equation involves infinite numbers of contributions in both channels.  
As a result, a direct analysis of the crossing equation is more challenging than those in the $\phi^{2n}$ cases.

In fact, the combination of the two methods leads to a relatively simple derivation of the leading term of $\hat\g_\F$ in the $\e$ expansion. 
In more details, there are infinitely many unknown BOE coefficients in the boundary channel expansion of $\<\f(x)\f(y)\>$.
Most of them can be fixed using the multiplet recombination.
Then, $\<\f(x)\f(y)\>$ is reduced to a function with a finite number of unknown parameters, 
and thus can be directly expanded in the bulk OPE limit.
By matching with the bulk channel expansion, the unknown parameters in the boundary channel  are determined, 
which include the anomalous dimension and the BOE coefficient of $\F$.

It is straightforward to generalize the above discussions to higher-derivative theories, 
which involve a higher-derivative kinetic term $\f\Box^k\f$.
As concrete examples, we carry out the explicit computation for $k=2,3$.
As in the canonical $k=1$ case, a higher derivative generalization also has a conformal IR fixed point associated with the $\f^3$ interaction,\footnote{In comparison with the canonical case ($k=1$), the higher-derivative fixed points with $k>1$ are more likely to be unstable since there are usually more relevant deformations.} and can be studied using the same procedure described above.
The $k=2$ case might be more interesting because it is related to the isotropic Lifshitz point (see \cite{Diehl:2002ri} and references therein).\footnote{A $\f^4$ theory version of the Lifshitz point can be realized on the lattice by the axial-next-nearest-neighbor Ising (ANNNI) model \cite{PhysRevLett.44.1502}.}
To the best of our knowledge, the $k>1$ boundary results are derived here for the first time.
The main results of this work are summarized in table \ref{summary}.

\begin{table}
	\centering
	\begin{tabular}{|c|c|c|}
		\hline
		Kinetic term: $\phi \Box^k \phi$ & Yang-Lee: $(\h\g_{\F}, \m_{\F}^{}, a_{\f^3})$ & Potts: $(\h\g_{\F}, \m_{\F}^{}, a_{\f^3})$\\
		\hline 
		$k=1$ & \makecell{\eqref{YL boundary results gamma}, \eqref{YL boundary results mu}, \eqref{a phi3 YL}} & \makecell{\eqref{Potts boundary channel result gamma}, \eqref{Potts boundary channel result mu} \eqref{a phi3}}\\
		\hline
		$k=2$ & \makecell{\eqref{YL k=2 boundary results gamma}, \eqref{YL k=2 boundary results mu}, \eqref{a phi3 YL k=2}} & \makecell{\eqref{Potts k=2 boundary results gamma}, \eqref{Potts k=2 boundary results mu}, \eqref{a phi3 k=2}}\\
		\hline
		$k=3$ & \makecell{\eqref{YL k=3 boundary results gamma}, \eqref{YL k=3 boundary results mu}, \eqref{a phi3 YL k=3}} & \makecell{\eqref{Potts k=3 boundary results gamma}, \eqref{Potts k=3 boundary results mu}, \eqref{a phi3 k=3}}\\
		\hline
	\end{tabular}
	\caption{Summary of the main results $(\h\g_{\F}, \m_{\F}^{}, a_{\f^3})$ for different $\phi^3$ CFTs with a boundary.
	Here $\h\g_{\F}^{}$ is the anomalous dimension of the boundary fundamental operator, $\m_\F$ is the coefficient of the bulk-boundary two-point function $\<\f(x)\F(\hat y)\>=\frac {\m_\F} {|x-\h y|^{2\h\D_\F}(2x_\perp)^{\D_\f-\h\D_\F}}$, 
	and $a_{\cal O}$ is the bulk one-point function coefficient, i.e., $\<\cal O(x)\>=\frac{a_{\cal O}}{(2x_\perp)^{\D_{\cal O}}}$.
	Here $|x-\hat{y}|$ is the distance between $x$ and $\hat y$. 
	In addition, for the single-field $\phi^{2n+1}$ theory, the one-point coefficient $a_\f$ in \eqref{general a phi} is for general $(k,n)$. We assume that $k$ and $2n-1$ have no common divisor.}
	\label{summary}
\end{table}

This paper is organized as follows.
In Sec. \ref{Yang-Lee model with a boundary}, we consider the Yang-Lee model with a boundary. 
We first illustrate the procedure for deriving the anomalous dimension of the boundary fundamental operator using the canonical example, 
and then discuss the higher-derivative generalizations of the Yang-Lee BCFT.
In Sec. \ref{Potts model with a boundary}, we study the case of the Potts model, which is a straightforward extension of the Yang-Lee case.
In Sec. \ref{Towards phi2n+1 theories}, we discuss the attempt to extend the above procedure to the $\f^{2n+1}$ theories with $n>1$.

\section{Yang-Lee model with a boundary}
\label{Yang-Lee model with a boundary}

We start with the canonical Yang-Lee model with a boundary, which corresponds to the action
\be
S_{\pm}\propto\int_{\bb R^d_+} \rm d^d x\(\f\Box\f+g\m^{\e/2}\f^3\)+\s\int_{\bb R^{d-1}}\F^{(0)}\F^{(1)}+\ldots
\,,
\ee
where $\m$ has mass dimension 1, the ellipsis denotes higher order terms in $\e$, and $g\in i\bb R$ is a purely imaginary coupling constant.
Here $\s=\pm 1$ determines the boundary condition.
The sign $+$ ($-$) corresponds to the Neumann (Dirichlet) boundary condition. 
We use the uppercase forms of Greek letters to denote the boundary operators,
\be
\F^{(0)}(\h x)=\lim_{x_\perp\ra 0}\f(x)\,,\qquad
\F^{(1)}(\h x)=\lim_{x_\perp\ra 0}\pa_\perp\f(x)
\,,
\ee
where the superscript indicates the number of derivatives.
Our goal is to obtain the anomalous dimension of the boundary fundamental operator $\F$. 
We have $\F=\F^{(0)}$ in the Neumann case, while $\F=\F^{(1)}$ in the Dirichlet case.
Without specifying a boundary condition, we use
\be\label{q sigma relation}
q=\frac{1-\s}{2}
\ee
to denote the number of derivatives in the boundary fundamental operator $\F^{(q)}$.

Since the boundary channel expansion of $\<\f(x)\f(y)\>$ contains the $\F^{(q)}$ contribution, 
our strategy is to extract the anomalous dimension of $\F^{(q)}$ by solving this bulk two-point function.
Due to boundary conformal symmetry, the two-point function takes the form
\be\label{phi 2pt}
\<\f(x)\f(y)\>=\frac {F_1(\x)} {(4x_\perp y_\perp)^{\D_\f}}\,, \qquad
\ee
where the cross ratio is
\be
\x=\frac {|x-y|^2} {4x_\perp y_\perp}
\,,
\ee
and the subscript 1 in $F_1(\x)$ indicates the power of $\f$ in the external bulk operators on the LHS of \eqref{phi 2pt}. 
Before studying the $\phi^3$ deformation of $F_1(\x)$, let us first examine the free theory solution,
\be\label{free F1}
F_{1,\text f}(\x)=\frac{1}{\x^{\frac{d-2}{2}}}+\frac{\s}{(1+\x)^{\frac{d-2}{2}}}
\,.
\ee
In the boundary channel expansion of $\<\f(x)\f(y)\>_{\text f}$, only the boundary fundamental operator $\F^{(q)}_{\text f}$ contributes,
\be
F_{1,\text f}(\x)=\m_{\F^{(q)},\text f}^{2} \, f_{\text {bdy}}(\h\D_{\F^{(q)},\text f}\,,\xi)
\,.
\ee
We use $\m_{{\mathcal O}}$ to denote the coefficient of the boundary primary ${\mathcal O}$ in the BOE of $\f$.
The squared free BOE coefficient is
\be
\m_{\F^{(q)},\text f}^{2}=\frac{d+2}{4}-\s\frac{d-6}{4}
\,,
\ee
and the boundary channel scalar block is \cite{McAvity:1995zd}
\be\label{boundary block}
f_\text{bdy}(\h\D,\x)=\x^{-\h\D}\,{}_2F_1\!\(\h\D,\h\D-\frac{d}{2}+1;2\h\D-d+2;-\frac{1}{\x}\)
\,.
\ee
In the interacting fixed point at $d=6-\e$, the boundary channel expansion of $F_{1}(\x)$ involves three types of corrections:
\begin{enumerate}
\item New boundary primaries contribute to the boundary channel expansion.
\item The scaling dimension $\h\D_{\F^{(q)}}$ contains an anomalous dimension $\h\g_{\F^{(q)}}^{}=\h\D_{\F^{(q)}}-\h\D_{\F^{(q)},\text f}\,$.
\item The squared BOE coefficient $\m_{\F^{(q)}}^{2}$ receives a correction.
\end{enumerate}
We use the multiplet recombination constraints to fix the first type of correction.
Then we determine the remaining unknown parameters by the boundary crossing equation, 
which include the leading anomalous dimension $\h\g_{\F^{(q)}}^{}$ and the leading correction in $\m_{\F^{(q)}}^{2}$. 

\subsection{Multiplet recombination}
\label{Multiplet recombination 1}

The bulk multiplet recombination implies that the $\e\ra 0$ limit of  the descendant $\Box\f$ is identified with $\f^2_{\text f}$. To be more precise, we have
\be\label{recombination}
\lim_{\e\ra 0}\a^{-1}\Box\f=\f^2_{\text f}
\,.
\ee
The singular factor $\a^{-1}$ was computed earlier in the $\f^3$ theory without a boundary \cite{Guo:2024bll}.
But we do not need to refer to the previous result for $\a$ since it can be independently determined here.

According to \eqref{recombination}, 
we have some matching conditions for bulk-boundary two-point functions
\be\label{bulk boundary matching}
\lim_{\e\ra 0}\a^{-1}\<\Box\f(x)\Ps_{\h\D}(\h y)\>=\<\f^2(x)\Ps_{\h\D}(\h y)\>_{\text f}
\,,
\ee
where $\Ps_{\h\D}$ denotes the boundary composite primary with scaling dimension $\h\D$.
The matching condition \eqref{bulk boundary matching} constrains the BOE coefficients of $\Ps_{\h\D}$, 
which is useful for resuming the boundary channel expansion of $\<\f(x)\f(y)\>$ in Sec. \ref{Resummation of phiphi}.

Above we sketched an approach to obtain the new boundary channel contributions in $\<\f(x)\f(y)\>$.\footnote{By new contributions we mean the contributions from the boundary primaries that do not appear in the free BOE, i.e., the first type of correction discussed below \eqref{boundary block}.} 
Instead of resumming the boundary channel expansion, we also consider the differential equation approach in Sec. \ref{Differential equation for phiphi}. 
The matching condition for $\<\f(x)\f(y)\>$,
\be\label{bulk bulk 2pt matching}
\lim_{\e\ra 0}\a^{-2}\<\Box\f(x)\Box\f(y)\>=\<\f^2(x)\f^2(y)\>_{\text f}
\,,
\ee
implies a useful differential equation for $F_1(\x)$.
We can deduce the total contributions of new boundary operators directly from the solution to this differential equation.

\subsubsection{Resummation of $\<\f(x)\f(y)\>$}
\label{Resummation of phiphi}

The matching condition \eqref{bulk boundary matching} leads to
\be\label{bulk boundary matching explicit}
&\lim_{\e\ra 0}\[\frac {(\D_\f-\h\D)(\D_\f-\h\D+1)} {|2x_\perp|^2}+\frac {\h\D(\D_\f-\frac {d-2} {2})} {|x-\h y|^2}\] \frac {4\a^{-1}\, \m_{\h\D}} {|x-\h y|^{2\h\D}(2x_\perp)^{\D_\f-\h\D}} \nn
=\;&\frac {\n_{\h\D,\text f}} {|x-\h y|^{2\h\D_{\text f}} (2x_\perp)^{2\D_{\f,\text f}-\h\D_{\text f}}}
\,.
\ee
Here we define $\m_{\h\D}\equiv \m_{\Ps_{\h\D}}^{}$ for brevity, and similarly $\n_{\h\D}\equiv \n_{\Ps_{\h\D}}^{}$ is the coefficient of $\Ps_{\h\D}$ in the BOE of $\f^2$.
The scaling dimension of $\f$ is
\be\label{bulk phi gamma}
\D_\f=\frac{d-2} 2+\g_\f
\,.
\ee
Since $\g_\f$ is of order $\a^2$ \cite{Guo:2023qtt}, the second term in the square bracket in \eqref{bulk boundary matching explicit} is zero in the $\e\ra 0$ limit.
We find the relation
\be\label{mu nu relation}
\m_{\h\D}=\frac {\a\, \n_{\h\D,\text f}} {4(\h\D-2)(\h\D-3)}+O(\e)
\,.
\ee
The free BOE coefficients $\n_{\h\D,\text f}$ can be derived from
\be
\<\f^2(x)\f^2(y)\>_{\text f}=\frac {F_{2,\text{f}}(\x)} {(4x_\perp y_\perp)^{2\D_{\f, \text f}}}\,,
\ee
where the explicit expression at $d=6$ from Wick contractions is given by
\be\label{F2}
F_{2,\text{f}}(\x)=1+2\[\,\frac 1 {\x^2}+\frac {\s} {(1+\x)^2}\]^2
\,.
\ee
The boundary-block decomposition reads
\be
F_{2,\text{f}}(\x)=1+4(1+\s)f_{\text{bdy}}(4,\x)+8 f_{\text{bdy}}(6,\x)+\frac{10}{3} f_{\text{bdy}}(8,\x)+\ldots\,,
\ee
where  the boundary scalar block is defined in  \eqref{boundary block} and we have used the identity $\s^2=1$.
In the boundary block expansion, the scaling dimensions of the free internal boundary operators are non-negative even integers other than $2$.
They are even because the internal boundary operators are constructed from two identical boundary fundamental operators and contracted derivatives.
The corresponding coefficients of the boundary blocks read
\be\label{nu}
\n^2_{\h\D,\text f}=\begin{cases}
		\; 1 &\quad \h\D=0\\
		\; 4(1+\s) &\quad \h\D=4\\
		\; \frac{(\h\D-1)!(\h\D-2)!}{3(2\h\D-7)!} &\quad \h\D\geq 6 \,,
	\end{cases}
\ee
where $\h\D$ is an even integer.\footnote{In a slight abuse of notation, here $\h\D$ is the scaling dimension of $\Ps_{\h\D}$ at order $\e^0$.
This ambiguity does not matter as the higher order terms are irrelevant to our analyses.}
According to \eqref{mu nu relation}, we obtain
\be\label{mu result}
\m_{\h\D}^2=\(\frac {\a} {4(\h\D-2)(\h\D-3)}\)^2 \n_{\h\D,\text f}^2+O(\e^{3/2})
\,.
\ee
For large enough $\h\D$, the coefficient $\m_{\h\D}^2$ should correspond to a sum of degenerate contributions.
We do not need to unmix them due to the following.
Each squared BOE coefficients satisfies \eqref{mu nu relation}, and thus the sum of squared BOE coefficients also obey \eqref{mu nu relation}.
In this way, we obtain the total coefficient of each boundary block in the boundary channel expansion of $\<\f(x) \f(y)\>$.
We remark that $\m_0$ can also be obtained from the one-point function matching condition
\be
\lim_{\e\ra 0}\a^{-1}\<\Box\f(x)\>=\<\f^2(x)\>_{\text f}
\,.
\ee
From this condition, we derive
\be\label{1pt matching explicit}
\lim_{\e\ra 0}\frac {\a^{-1} \D_\f(\D_\f+3-d) \m_0} {2^{\D_\f}|x_\perp|^{\D_\f+2}}=\frac {\s} {(2x_\perp)^{2\D_{\f,\text f}}}
\,,
\ee
which yields the one-point function coefficient of $\f$
\be\label{a}
a_\f\equiv \m_0=\frac {\s} {24}\a+O(\e)
\,.
\ee
The subscript 0 corresponds to the dimension of the boundary identity.
Since $\s^2=1$, the squared one-point function coefficient is 
\be
\m_0^2=\frac{\a^2}{576}+O(\e^{3/2})
\,,
\ee
which is consistent with the $\h\D=0$ case of \eqref{mu result}.

Now we revisit the boundary-block expansion
\be\label{YL boundary channel}
F_1(\x)=\m_{\F^{(q)}}^{2} \, f_{\text {bdy}}(\h\D_{\F^{(q)}}\,,\xi)+\sum_{\h\D=0,4,6,8,\ldots}\m_{\h\D}^2 \, f_{\text {bdy}}(\h\D\,,\xi)
\,.
\ee
Using \eqref{nu} and \eqref{mu result}, the infinite sum of the new contributions can be resumed and the result reads
\be\label{F function result}
F_1(\x)=\m_{\F^{(q)}}^{2} \, f_{\text {bdy}}(\h\D_{\F^{(q)}}\,,\xi)+\(\frac 1 {576}+\frac{(1+2\x)\log\frac{1+\x}{\x}+1+3\s}{48\x^2(1+\x)^2}\)\a^2+O(\e^{3/2})
\,.
\ee
Below we discuss an alternative approach that does not involve the explicit resummation but leads to the same result \eqref{F function result}.

\subsubsection{Differential equation for $\<\f(x)\f(y)\>$}
\label{Differential equation for phiphi}

Let us examine the recombination constraint \eqref{bulk bulk 2pt matching} to order $\e^1$. 
The RHS of \eqref{bulk bulk 2pt matching} is given by \eqref{F2}.
The LHS of \eqref{bulk bulk 2pt matching} takes the form
\be
\<\Box\f(x)\Box\f(y)\>=\frac {\cal D F_1(\x)} {(4x_\perp y_\perp)^{\D_\f+2}}
\,,
\ee
where the differential operator is
\be\label{D}
\cal D=\;&16\x^2(1+\x)^2 \frac {\rm d^4} {\rm d \x^4}
+8\x(1+\x)\[4(\D_\f+2)(1+2\x)-2(\D_\f-\tfrac{d-2} 2) \frac {x_\perp^2+y_\perp^2} {x_\perp y_\perp} \] \frac {\rm d^3} {\rm d \x^3}\nn
& 4\Bigg[d^2-2 d (2 \D _\f+3)+8 (\x^2 (3\D_\f (\D_\f+3)+7)+3 \x  \D_\f (\D_\f+3)+\D_\f (\D_\f+3)+7\x +2) \nn
&\hspace{3mm} -2(\D_\f+2)(\D_\f-\tfrac{d-2} 2)(1+2\x) \frac {x_\perp^2+y_\perp^2} {x_\perp y_\perp}\Bigg]\frac {\rm d^2} {\rm d \x^2}\nn
&+8(\D_\f+1)\[4(\D_\f+1)^2(1+2\x)-2(\D_\f+2)(\D_\f-\tfrac{d-2} 2) \frac {x_\perp^2+y_\perp^2} {x_\perp y_\perp}\]\frac {\rm d} {\rm d \x}+16\D_\f^2(\D_\f+1)^2
\,.
\ee
The differential equation is
\be
\cal D F_1(\x)=\a^2 F_{2, \text f}(\x)+O(\e^{3/2})
\,.
\ee
To study the leading correction to the free theory solution, we write
\be\label{split F1}
F_1(\x)=F_{1, \text f} (\x)+\d F_1(\x)
\,,
\ee
where the free part is
\be\label{F1 free}
F_{1, \text f} (\x)=\frac{1}{\x^2}+\frac{\s}{(1+\x)^2}+\frac{1}{2}\[\frac{\log\x}{\x^2}+\s\frac{\log(1+\x)}{(1+\x)^2}\]\e+O(\e^2)
\,.
\ee
This is the expression \eqref{free F1} with $d=6-\e$.
One can check that $\cal D F_{1, \text f} (\x)=O(\e^{2})$.
For the leading correction, we have
\be
&16\Big[\x^2 (1+\x)^2 \frac {\rm d^4} {\rm d \x^4}\d F_1(\x)
+8 \x(1+\x)(1+2\x) \frac {\rm d^3} {\rm d \x^3}\d F_1(\x)
+2 (6+37 \x+37 \x^2) \frac {\rm d^2} {\rm d \x^2}\d F_1(\x)\nn
&+54 (1+2\x) \frac {\rm d} {\rm d \x}\d F_1(\x)
+36 \d F_1(\x)\Big]
=\a^2 F_{2,\text f}(\x)+O(\e^{3/2})
\,.
\ee
The solution is
\be\label{delta F result}
\d F_1(\x)=\;&\frac{c_1}{\x^2}+\frac{c_2}{(1+\x)^2}+c_3\[\frac{\log(1+\x)}{\x^2}-\frac{1+4\x}{2\x^2(1+\x)^2}\]+c_4\[\frac{\log\x}{(1+\x)^2}-\frac{1+3\x}{\x^2(1+\x)^2}\]\nn
&+\(\frac 1 {576}+\frac{(1+2\x)\log\frac{1+\x}{\x}+1+3\s}{48\x^2(1+\x)^2}\)\a^2+O(\e^{3/2})
\,,
\ee
where we have shifted the free parameters $c_i$ appropriately for convenience.
The boundary channel expansion implies that 
\be
c_1&=\frac{2(\m_{\F^{(q)}}^2)^{(1)}-(3+7\s)\h\g_{\F^{(q)}}^{(1)}}{4}\e+O(\e^{3/2})\,, \quad
c_2=\frac{2\s(\m_{\F^{(q)}}^2)^{(1)}+(3+7\s)\h\g_{\F^{(q)}}^{(1)}}{4}\e+O(\e^{3/2})\,, \nn
c_3&=-\h\g_{\F^{(q)}}^{(1)}\e+O(\e^{3/2})\,, \hspace{37.5mm}
c_4=-\s\h\g_{\F^{(q)}}^{(1)}\e+O(\e^{3/2})\,,
\ee
where  $(\m_{\F^{(q)}}^2)^{(1)}$ and $\h\g_{\F^{(q)}}^{(1)}$ are defined by
\be
\m_{\F^{(q)}}^2=\m_{\F^{(q)},\text f}^2+(\m_{\F^{(q)}}^2)^{(1)}\e+O(\e^{3/2})\,, \qquad
\h\D_{\F^{(q)}}=\h\D_{\F^{(q)},\text f}+\h\g_{\F^{(q)}}^{(1)}\e+O(\e^{3/2})
\,.
\ee
The terms with free parameters in the general solution \eqref{delta F result} can be absorbed into the leading correction to the $\F$ contribution.
In other words, we have
\be
\text{first line of \eqref{delta F result}}=\[(\m_{\F^{(q)}}^2)^{(1)} f_{\text {bdy}}(\h\D_{\F^{(q)},\text f}\,,\xi)+\m_{\F^{(q)},\text f}^2 \, \h\g_{\F^{(q)}}^{(1)}\frac{\pa}{\pa \h\D} f_{\text {bdy}}(\h\D\,,\xi)\Big|_{\h\D=\h\D_{\F^{(q)}}}\]\e+O(\e^{3/2})
\,,
\ee
so we obtain the same $F_1(\x)$ as the resummation result \eqref{F function result}.
In the next step, we impose boundary crossing symmetry on $F_1(\x)$ to fix the BCFT data.

\subsection{Boundary crossing equation}
\label{Boundary crossing equation}

There are two unknown parameters in \eqref{F function result}. 
Let us fix them by boundary crossing symmetry. 
In the bulk channel, we impose that \eqref{F function result} can be expanded in terms of bulk channel scalar blocks \cite{McAvity:1995zd}
\be
f_\text{bulk}(\D,\x)=\x^{\D/2}\,{}_2F_1\!\(\frac{\D}{2},\frac{\D}{2};\D-\frac{d}{2}+1;-\x\)
\,.
\ee
The bulk channel expansion reads
\be\label{bulk expansion YL}
\x^{\D_\f}F_1(\x)=1+\l_\f f_\text{bulk}(\D_\f,\x)+\sum_{\D=6}^{\infty}\l_{\D} f_\text{bulk}(\D,\x)+O(\e^{3/2})
\,.
\ee
Throughout this paper, we use $\l$ to denote the expansion coefficients in the bulk channel.
In other words, $\l$ represents the product of certain one-point function coefficient and bulk OPE coefficient.
Note that $\f^2$ becomes a descendant in the interacting Yang-Lee theory, and its contribution is absorbed into the $\f$ block. 
More specifically, in the $\e\ra 0$ limit the $\f$ block diverges as
\be
f_\text{bulk}(\D_\f,\x)=-\frac{1}{\g_\f}\frac{\x^2}{(1+\x)^2}+O(\e^0)
\,.
\ee
But the formal product $\l_\f f_\text{bulk}(\D_\f,\x)$ is regular because $\l_\f$ and $\g_\f$ are both of order $\e$.
In the $\e\ra 0$ limit, it reduces to the $\f^2_{\text f}$ contribution,
\be
\lim_{\e\ra 0}\l_\f f_\text{bulk}(\D_\f,\x)=\l_{\f^2,\text f}f_\text{bulk}(\D_{\f^2,\text f},\x)
\,.
\ee
We expand \eqref{bulk expansion YL} in $\x$ and solve it order by order.
The parts with and without $\log\x$ yield independent constraints.
For low orders in $\x$,
we have assumed that the bulk identity contribution does not receive a correction and there is no new contribution with $\D<\D_\f$.
These assumptions determine $\h\g_{\F^{(q)}}^{}$, $\m^2_{\F^{(q)}}$, and $\g_\f$ in terms of $\a^2$.
Moving on to relatively higher order terms in $\x$, we take into account the $\f$ contribution and fix the parameters $\a^2$ and $\l_\f$.
Finally, $\l_\D$ with $\D \geq 6$ are determined by examining more higher order terms in $\x$.
Below, we report the results that follow from \eqref{bulk expansion YL}.
The parameters in the boundary channel decomposition \eqref{F function result} are given by
\be
\a^2&=-\frac{8\e}{3} +O(\e)\,,\\
\h\g_{\F^{(q)}}^{}&=-\frac {5+ \s}{18}\e+O(\e^{3/2})\,, \label{YL boundary results gamma}\\
\m^2_{\F^{(q)}}&=\[2+\frac{(\s-1)\e}{4}\]-\frac{(1+\s)\e}{9}+O(\e^{3/2}) \label{YL boundary results mu}
\,,
\ee
where the terms in the square bracket correspond to the free-theory solution.
In the Dirichlet case, the surface critical exponent in the special transition,
\footnote{In the Yang-Lee model with a boundary, we need to take into account the relevant boundary operator $\Phi$, so the Dirichlet case is associated with a multicritical point.}
\be
\eta_\parallel=2\h\g_{\F^{(1)}}^{}=-\frac{4}{9}\e+O(\e^{3/2})\,,
\ee
agrees with the result in \cite{Janssen1995}.\footnote{The anomalous dimension of the boundary fundamental operator is also obtained in \cite{Dey:2020jlc}, but their result is different from ours.
It seems that there might be some typos in \cite{Dey:2020jlc}.}
The results for the bulk parameters in \eqref{bulk expansion YL} are\footnote{We write the higher order terms of the bulk anomalous dimension $\g_\f$ as $O(\e^2)$ since it is known to be of integer order in $\e$ (see e.g. \cite{Borinsky:2021jdb}).}
\be\label{bulk result}
\g_\f&=-\frac{\e}{18}+O(\e^2)\,,\nn
\l_\f&=\frac{\s}{18}\e+\frac{1}{216}\(\frac{47}{18}+5\s\)\e^2+O(\e^{5/2})\,,\nn
\l_{\f^3}&=\frac{1}{54}\(\frac{13}{2}+\s\)\e+O(\e^{3/2})\,,
\ee
and for $\D\geq 8$,
\be\label{bulk result 2}
\l_{\D}=-\frac{((\D-4)(\D-2)+48(-1)^{\D/2})\G(\tfrac{\D}{2})\G(\tfrac{\D-4}{2})}{864(\D-4)!}\e+O(\e^{3/2}) 
\,,
\ee
where $\D$ is an even integer.
We are able to derive the subleading correction to $\l_\f$ due to the multiplet recombination at the level of bulk channel conformal blocks.
The bulk anomalous dimension $\g_\f$ is in agreement with \cite{Guo:2024bll}.
Using \eqref{a}, the ratio $\l_\f/\m_0$ gives the OPE coefficient of $\f\in\f\times\f$
\be
\l_\f/\m_0=\pm\sqrt{-\frac{2\e}{3}}+O(\e)
\,,
\ee
which also agrees with our previous bulk determination in \cite{Guo:2024bll}.
The one-point function coefficient of $\f^3$ is given by
\be\label{a phi3 YL}
a_{\f^3}=\l_{\f^3}/C_{\f^3}=\pm\frac{1}{54}\sqrt{-\frac{3}{2}}\(\frac{13}{2}+\s\)\e^{1/2}+O(\e)
\,,
\ee
where the bulk OPE coefficient $C_{\f^3}$ was derived in appendix F of \cite{Guo:2024bll}.

\subsection{Higher derivative generalizations}
\label{Higher derivative generalizations 1}

The discussion above can be generalized straightforwardly to higher derivative theories.
For the $\Box^k$ theory, the upper critical dimension is $d_{\text u}=6k$.
Analogous to \eqref{bulk boundary matching}, the matching condition for bulk-boundary two point functions is
\be\label{general k bulk boundary matching}
\lim_{\e\ra 0}\a^{-1}\<\Box^k \f(x)\Ps_{\h\D}(\h y)\>=\<\f^2(x)\Ps_{\h\D}(\h y)\>_{\text f}
\,,
\ee
which leads to
\be\label{boxk 2pt}
\lim_{\e\ra 0}\Bigg[\a^{-1}\sum_{l=0}^k\Bigg(\;\frac{(-1)^l 4^{k}k!}{(k-l)!\,l!}\frac{(\D_\f-\h\D)_{2k-2l}\,(\h\D)_l\,(-\g_\f)_l\,\m_{\h\D}}{|x-y|^{\h\D+l}(2x_\perp)^{\D_\f-\h\D+2k-2l}}\,\Bigg)\Bigg]
=\frac{\n_{\h\D,\text f}}{|x-y|^{\h\D}(2x_\perp)^{2n\D_{\f,\text f}-\h\D_{\text f}}}
\,.
\ee
Here $(a)_b=\G(a+b)/\G(a)$ is the Pochhammer symbol.

As a concrete example, we will focus on the $k=2$ case. 
The results for the $k=3$ case can also be found in Appendix \ref{Results for the k=3 case}.
As a generalization of the $\s$ in the $k=1$ case, we define 
\be
\s_i=\begin{cases}
	\; +1 \quad & \text{(N)} \\
	\; -1  & \text{(D)}\,,
\end{cases}
\ee
where $i=0,1$ for $k=2$, and N (D) indicates the boundary condition of the Neumann (Dirichlet) type.
The set $\{\s_0,\s_1\}$ labels the boundary conditions.
For different choices of $\{\s_0,\s_1\}$, the boundary fundamental operators are different.
It is convenient to introduce another set of variables $\{q_0,q_1\}$, which corresponds to the number of transverse derivatives in the boundary fundamental operators.
The two sets of variables are related by
\be
q_i=i\frac{1+\s_i}{2}+(3-i)\frac{1-\s_i}{2}
\,,
\ee
which is the $k=2$ generalization of \eqref{q sigma relation}.
The matching condition \eqref{boxk 2pt} yields
\be\label{k=2 mu}
\m_{\h\D}^2=\(\frac {\a} {16(\h\D-4)(\h\D-5)(\h\D-6)(\h\D-7)}\)^2 \n_{\h\D,\text f}^2+O(\e^{3/2})
\,.
\ee
Next, we compute the squared free BOE coefficient $\n_{\h\D,\text f}^2$.
Using Wick contractions, we find
\be
F_{2,\text f}(\x)=(3\s_0+2\s_1)^2+\frac{2\(1+5\x+10\x^2+10\x^3+(5+3\s_0+2\s_1)\x^4+(1+\s_0)\x^5\)^2}{\x^8 (1+\x)^{10}}
\,.
\ee
The boundary channel expansion of $F_{2,\text f}(\x)$ yields
\be\label{k=2 nu}
\n_{\h\D,\text f}^2&=\begin{cases}
		\; 13+12\s_0\s_1&\quad \h\D=0\\
		\; 4(1+\s_0) &\quad \h\D=8\\
		\; 8(1+\s_0)(1+\s_1) &\quad \h\D=9\\
		\; \frac{4}{7}(19-9\s_0-7\s_1-35\s_0\s_1) &\quad \h\D=10
		\,,
	\end{cases}
\ee
and for $\h\D\geq 12$,
\be\label{k=2 nu high Delta}
\n_{\h\D,\text f}^2=\;&\frac{(\h\D-7)(\h\D-6)(\h\D-4)!(\h\D-1)!}{544320(2\h\D-12)!}\nn
&\times[108(1+(-1)^{\h\D})+(-1)^{\h\D}(\h\D-11)(\h\D-8)(\h\D-3)\h\D(1+\s_0\s_1)]
\,.
\ee
Here $\h\D$ is an integer.
We have used the identity $\s_i^2=1$ for all $i$.
The coefficient $\n_{\h\D,\text f}^2$ vanishes for other $\h\D$.
As opposed to \eqref{nu}, the BOE coefficients $\n_{\h\D,\text f}$ can be nonzero for odd $\h\D$ if $\s_0\s_1\neq -1$.
This is because the internal boundary primaries can be constructed from two different boundary fundamental operators.
If $\s_0\s_1\neq -1$, then the sum of the two corresponding scaling dimensions is odd.
The boundary channel resummation leads to
\be
F_1(\x)=\;&\sum_{i=0}^1\m_{\F^{(q_i)}}^2 f_{\text {bdy}}(\h\D_{\F^{(q_i)}},\x)+\frac {\a^2} {4877107200 \xi ^5 (\xi +1)^6}
\Bigg\{(27 \xi ^{11}+162 \xi ^{10}+405 \xi ^9 +540 \xi ^8 \nn
&+405 \xi ^7)(12 \s_0 \s_1+13)-6 \xi ^6 (3596 \s_0 \s_1+3569)+\xi ^5 (84351-6396 \s_0 \s_1)\nn
&+280 \xi ^4 (23 \s_0\s_1+995)-140 \xi ^3 (5 \s_0 (4 \s_1-189)-3247)+84 \xi ^2 (7 (18\s_1+805)\nn
&+\s_0 (145 \s_1+3276))+28 \xi  (546 \s_1+5
\s_0 (85 \s_1+1323)+8660)\nn
&+56 (84 \s_1+\s_0 (55 \s_1+756)+865)+840 (\xi +1) \xi  \Big[28 \xi ^5
(\s_0 \s_1+1)-6 \xi ^4 (\s_0 \s_1+19)\nn
&-270 \xi ^3-270 \xi^2-135 \xi -27\Big] \log\frac{1+\x}{\xi }\Bigg\}+O(\e^{3/2})
\,.
\ee
In the bulk channel expansion, we expect to have
\be
\x^{\D_\f}F_1(\x)=1+\l_\f f_\text{bulk}(\D_\f,\x)+\l_{\pa^2\f^2} f_\text{bulk}(\D_{\pa^2\f^2},\x)+\sum_{\D=12}^{\infty}\l_{\D} f_\text{bulk}(\D,\x)+O(\e^{3/2})
\,.
\ee
A difference from \eqref{bulk expansion YL} is that now $\pa^2\f^2$ contributes.
The consistency of $F_1(\x)$ with the bulk channel expansion determines the unknown parameters. 
We obtain
\be
\a^2&=-\frac{322560}{37}\e+O(\e^{3/2})\,,\\
\h\g^{}_{\F^{(q_0)}}&=\(\frac {45} {74}+\frac {55} {666}\s_0+\frac {14} {333}\s_1+\frac {31} {111}\s_0\s_1\)\e+O(\e^{3/2})\,, \nn
\h\g^{}_{\F^{(q_1)}}&=-\(\frac {24} {37}+\frac {5} {74}\s_0+\frac {4} {37}\s_1+\frac {93} {74}\s_0\s_1\)\e+O(\e^{3/2})\,, \label{YL k=2 boundary results gamma}\\
\m^{}_{\F^{(q_0)}}&=\[6\s_0-4+\frac{37(1-\s_0)}{24}\e\]+\(\frac{171673}{19980}-\frac{115957\s_0}{19980}-\frac{53557\s_1}{19980}+\frac{83593\s_0\s_1}{19980}\)\e+O(\e^{3/2})\,,\nn
\m^{}_{\F^{(q_1)}}&=\[7-3\s_1+\frac{7\s_1-11}{8}\e\]+\(-\frac{952}{185}+\frac{6761\s_0}{1110}+\frac{1264\s_1}{555}-\frac{4053\s_0\s_1}{370}\)\e+O(\e^{3/2}) \label{YL k=2 boundary results mu}
\,.
\ee
Again, the terms in the square brackets corresponds to the free-theory solutions.
The bulk channel results are
\be\label{YL k=2 bulk results}
\g_\f&=\frac{3}{74}\e+O(\e^{3/2})\,, \qquad 
\g_{\pa^2\f^2}=-\frac{11}{37}\e+O(\e^{3/2})\,, \nn
\l_\f&=-\(\frac{\s_0}{148}+\frac{\s_1}{222}\)\e-\(\frac{21311}{1839936}+\frac{11651 \sigma _0}{1971360}+\frac{1249\sigma _1}{328560}+\frac{16157 \sigma _0 \sigma _1}{1379952}\)\e^2+O(\e^{5/2})\,, \nn
\l_{\pa^2\f^2}&=2\s_0+\frac{2\s_1}{3}+\(\frac{6851}{3108}+\frac{7589 \sigma _0}{6660}+\frac{1349 \sigma_1}{2220}+\frac{110603 \sigma _0 \sigma _1}{46620}\)\e+O(\e^{3/2})\,, \nn
\l_{\f^3}&=-\(\frac{487}{2072}+\frac{65 \sigma _0}{3108}+\frac{11 \sigma_1}{777}+\frac{93\sigma _0 \sigma _1}{518}\)\e+O(\e^{3/2})\,,\nn
\l_{\pa^2\f^3}&=\(\frac{1171}{43512}+\frac{25 \sigma _0}{32634}+\frac{25 \sigma _1}{32634}+\frac{363
   \sigma _0 \sigma _1}{14504}\)\e+O(\e^{3/2})
\,,
\ee
and the coefficients with $\D\geq 16$ are
\be\label{YL k=2 bulk results 2}
\l_{\D}&=-\frac{\sqrt{\pi } (\Delta -6)^2 (\Delta -4) (\Delta -2)\Gamma (\frac{\Delta
   }{2}-5)
}{333\times 2^{\Delta+5} \Gamma (\frac{\Delta -5}{2})}
   \Big[\frac{4}{35} (1-\D/2)_5\,
   (12 \s_0\s_1 +13)\nn
   &+\frac{1}{3} (-1)^{\Delta/2} (\D(\D-12)(\D(\D-12)+188) (\s_0\s_1  +1)+576 (8 \s_0\s_1 +11))\Big]\e+O(\e^{3/2})
\,.
\ee
The bulk anomalous dimensions $\g_\f$ and $\g_{\pa^2\f^2}$ agree with the previous results in \cite{Guo:2024bll}.
As in the $k=1$ result \eqref{bulk result}, the subleading correction to $\l_\f$ is obtained due to the multiplet recombination at the level of bulk channel conformal blocks.
We also determine the one-point function coefficient of $\f^3$
\be\label{a phi3 YL k=2}
a_{\f^3}=\l_{\f^3}/C_{\f^3}=\pm3\sqrt{-\frac{37}{35}}\(\frac{487}{2072}+\frac{65 \sigma _0}{3108}+\frac{11 \sigma_1}{777}+\frac{93\sigma _0 \sigma _1}{518}\)\e^{1/2}+O(\e)
\,.
\ee

\section{Potts model with a boundary}
\label{Potts model with a boundary}

The action for the Potts model with a boundary is
\be
S_{\pm}\propto\int_{\bb R^d_+} \rm d^d x\Big(\sum_{a=1}^N\f_a\Box\f_a+g\m^{\e/2} \sum_{a,b,c=1}^N T_{abc}\f_a\f_b\f_c\Big)+\s\int_{\bb R^{d-1}}\sum_{a=1}^N\F^{(0,1)}_a\F^{(1,1)}_a+\ldots
\,,
\ee
where $\f_a$ with $a=1,\ldots\,, N$ transforms in the $S_{N+1}$ standard representation, 
and $T_{abc}$ is the rank-3 $S_{N+1}$-invariant tensor.\footnote{Following \cite{Zia:1975ha,Guo:2024bll}, the tensor $T_{abc}$ is defined as
\be\label{T-definition}
	T_{abc}=\sum_{\a=1}^{N+1} e_{a}^{\a} e_{b}^{\a} e_{c}^{\a}
	\,.
\ee
Here $\{e_a^1,\ldots,e_a^{N+1}\}$ is a set of vectors pointing from the center to the vertices of an $N$-simplex.
They satisfy the conditions
\be\label{e-relation}
	\sum_{\a=1}^{N+1}e_{a}^{\a}=0
	\,, \qquad
	\sum_{\a=1}^{N+1}e_{a}^{\a}e_{b}^{\a}=(N+1)\d_{ab}
	\,, \qquad
	\sum_{a=1}^{N}e_{a}^{\a}e_{a}^{\b}=(N+1)\d^{\a\b}-1
	\,.
\ee
See Appendix C of \cite{Guo:2024bll} for more discussion on our convention.}
As in Sec. \ref{Yang-Lee model with a boundary}, the goal is to determine the anomalous dimensions of boundary fundamental operators, and we can proceed with the same strategy.

Analogous to \eqref{recombination}, the multiplet recombination is associated with
\be
\lim_{\e\ra 0}\a^{-1}\Box\f_a=\sum_{b,c=1}^N T_{abc}\,\f_{b,\text f}\,\f_{c,\text f}
\,,
\ee
and $\a$ is given by \cite{Guo:2024bll}
\be\label{Potts alpha}
\a^2=-\frac {8\e} {(3N-7)(N+1)^2}+O(\e^{3/2})
\,.
\ee
Below, we will use the shorthand notation
\be\label{K-Tphiphi}
K_{a,\text f}\equiv \sum_{b,c=1}^N T_{abc}\f_{b,\text f}\f_{c,\text f}
\,.
\ee
In contrast to the Yang-Lee case, $\a$ will not be independently determined here, so we will use \eqref{Potts alpha} as an input.

The bulk two-point function involves a tensor product of $S_{N+1}$ standard representations, so we quote the result in \cite{murnaghan1938analysis} for the tensor product decomposition. 
For positive integers $N>2$, the tensor product decomposition reads
\begin{align}\label{tensor-decomposition}
	(N,1)\otimes(N,1)=(N+1)+(N,1)+(N-1,2)+(N-1,1^{2})
	\,,
\end{align}
where the corresponding Young diagrams are
\begin{align}
	(N+1)&=
	\overbrace{
		\begin{ytableau}
			{} & {} & \none[\ldots] & {}
	\end{ytableau} }^{N+1 \text{ columns} }
	\,,
	\qquad\qquad\quad\;\;
	(N,1)=
	\overbrace{
		\begin{ytableau}
			{} & {} & \none[\ldots] & {} \\
			{}
	\end{ytableau} }^{N \text{ columns} }
	\,,
	\nn
	{}
	\nn
	(N-1,2)&=
	\overbrace{
		\begin{ytableau}
			{} & {} & {} & \none[\ldots] & {} \\
			{} & {}
	\end{ytableau} }^{N-1 \text{ columns} }
	\,,
	\qquad
	(N-1,1^{2})=
	\overbrace{
		\begin{ytableau}
			{} & {} & \none[\ldots] & {} \\
			{} \\
			{}
	\end{ytableau} }^{N-1 \text{ columns} }
	\,.
\end{align} 
The irreducible representations $(N+1)$ and $(N,1)$ are referred to as the trivial and standard representations, respectively. 
They appear in the crossing analysis of the bulk two-point function.

\subsection{Multiplet recombination}

The one-point function of the bulk fundamental scalar is non-zero in the Yang-Lee case, 
but it vanishes here due to unbroken $S_{N+1}$ symmetry.\footnote{In the Yang-Lee model, assuming that $\f\ra-\f$ under space reflection $\cal P$ and $i\ra -i$ under time reversal $\cal T$, the unbroken $\cal {PT}$ symmetry does not require $\<\f\>$ to vanish.
We only know that the one-point function $\<\f\>$ is purely imaginary.}
One can confirm this by considering the matching condition\footnote{In contrast to \eqref{bulk boundary matching}, we do not treat \eqref{1pt matching Potts} as the $\h\D=0$ case of \eqref{bulk boundary 2pt matching}, since the boundary identity and $\big(\Ps_{\h\D}\big)_b$ are in different irreducible representations of $S_{N+1}$.}
\be\label{1pt matching Potts}
\lim_{\e\ra 0}\a^{-1}\<\Box\f_a(x)\>=\<K_a(x)\>_{\text f}=0
\,.
\ee
The free one-point function on the RHS is zero because 
$\<K_a(x)\>_{\text f}\propto \sum_{a,b=1}^{N}T_{abc}\,\d_{bc}=0$.
Since $\a$ is of order $\e^{1/2}$, this implies that one-point function coefficient of $\f$ vanishes to order $\e^{1/2}$.
Similar to \eqref{bulk boundary matching}, we consider the matching conditions for bulk-boundary two-point functions,
\be\label{bulk boundary 2pt matching}
\lim_{\e\ra 0}\a^{-1}\<\Box\f_a(x)\big(\Ps_{\h\D}\big)_b(\h y)\>=\<K_a(x)\big(\Ps_{\h\D}\big)_b(\h y)\>_{\text f}
\,.
\ee
This gives the same equation as \eqref{bulk boundary matching explicit}, except that an overall tensor structure $\d_{ab}$ will appear on both sides in the Potts case.

\subsubsection{Resummation of $\<\f(x)\f(y)\>$}

The free BOE coefficient on the RHS of \eqref{bulk boundary 2pt matching} is computed by the boundary channel expansion of
\be
\<K_a(x)K_b(y)\>_{\text f}=\frac {\d_{ab}F_{2,\text f}(\x)} {(4x_\perp y_\perp)^{2\D_{K, \text f}}}
\,.
\ee
In the $d=6$ free theory, the function $F_{2,\text f}(\x)$ is obtained using Wick contractions,
\be\label{F2}
F_{2,\text f}(\x)=\frac {2(N-1)(N+1)^2((1+\x)^2+\s\x^2)^2} {\x^4(1+\x)^4}
\,,
\ee
where we have used $\sum_{c,e=1}^N T_{ace}T_{bce}=(N-1)(N+1)^2 \d_{ab}$.
Expanding $F_{2,\text f}(\x)$ in terms of the boundary blocks \eqref{boundary block}, we find the squared free BOE coefficients
\be
\n^2_{\h\D,\text f}=\begin{cases}
		\;  4(1+\s)(N-1)(N+1)^2 &\quad \h\D=4\\
		\; \frac{(\h\D-1)!(\h\D-2)!}{3(2\h\D-7)!}(N-1)(N+1)^2 &\quad \h\D\geq 6 \,,
	\end{cases}
\ee
which are $(N-1)(N+1)^2$ times those of \eqref{nu} except for the $\h\D=0$ case.
The relation \eqref{mu result} still holds, and it determines the squared BOE coefficients $\m_{\h\D}^2$.
As in the Yang-Lee case, we do not unmix the degenerate operators.

The boundary channel expansion of $F_1(\x)$ reads
\be
F_1(\x)=\m_{\F^{(q)}}^{2} \, f_{\text {bdy}}(\h\D_{\F^{(q)}}\,,\xi)+\sum_{\h\D=4,6,8,\ldots}\m_{\h\D}^2 \, f_{\text {bdy}}(\h\D\,,\xi)
\,.
\ee
This is similar to \eqref{YL boundary channel}, but the boundary identity does not contribute due to the vanishing one-point coefficient of $\phi_a$.
The resummation of boundary channel expansion leads to
\be\label{F function result Potts}
F_1(\x)=\m^2_{\F^{(q)}} f_{\text {bdy}}(\h\D_{\F^{(q)}},\x)+\frac {(N-1)(N+1)^2} {48} \(\frac{(1+2\x)\log(\frac{1+\x}{\x})+1+3\s}{\x^2(1+\x)^2}\) \a^2+O(\e^{3/2})
\,.
\ee
Some differences from \eqref{F function result} are that an $N$-dependent factor is present and the constant term is absent here.
Following the discussion in the Yang-Lee case, we also use the differential equation approach to obtain the same result as \eqref{F function result Potts}.

\subsubsection{Differential equation for $\<\f(x)\f(y)\>$}

The procedure in Sec. \ref{Differential equation for phiphi} can be generalized to the Potts model.
The bulk two-point function reads
\be
\<\Box\f_a(x)\Box\f_b(y)\>=\frac {\d_{ab}\cal D F_1(\x)} {(4x_\perp y_\perp)^{\D_\f}}
\,,
\ee
where $\cal D$ is given by \eqref{D}.
The differential equation is $\cal D F_1(\x)=\a^2 F_{2,\text f}(\x)+O(\e^{3/2})$, where $F_{2,\text f}(\x)$ is obtained in \eqref{F2}.
As in \eqref{split F1}, we single out the free part of $F_1(\x)$, which is again given by \eqref{F1 free}.
Then, the solution for $\d F_1(\x)$ is
\be
\d F_1(\x)=\;&\frac{c_1}{\x^2}+\frac{c_2}{(1+\x)^2}+c_3\[\frac{\log(1+\x)}{\x^2}-\frac{1+4\x}{2\x^2(1+\x)^2}\]+c_4\[\frac{\log\x}{(1+\x)^2}-\frac{1+3\x}{\x^2(1+\x)^2}\]\nn
&+\frac {(N-1)(N+1)^2} {48} \(\frac{(1+2\x)\log(\frac{1+\x}{\x})+1+3\s}{\x^2(1+\x)^2}\) \a^2+O(\e^{3/2})
\,,
\ee
where the first line takes the same form as that of \eqref{delta F result}. 
The second line is proportional to the second term in the second line of of \eqref{delta F result}.

\subsection{Boundary crossing equation}

To further determine the unknown parameters, we expand \eqref{F function result Potts} in the bulk channel,
\be
\x^{\D_\f}F_1(\x)=1+\l_{\f^2} f_\text{bulk}(\D_{\f^2},\x)+\sum_{\D=4}^{\infty}\l_{\D} f_\text{bulk}(\D,\x)+O(\e^{3/2})
\,,
\ee
where $\f^2$ is in the trivial representation of $S_{N+1}$.
There are two differences from \eqref{bulk expansion YL}.
Firstly, there is no $\f_a$ contribution because only operators in trivial representation can have nonzero one-point coefficient, otherwise the $S_{N+1}$ symmetry is broken.
Secondly, $\f^2$ is a primary, not a descendant, because $\Box \phi$ is associated with the bilinear operator $K_a$ in \eqref{K-Tphiphi}.
We find the boundary channel parameters
\be
\h\g_{\F^{(q)}}&=\frac{6+\s}{48}(N-1)(N+1)^2\a^2+O(\e^{3/2})\,, \label{Potts boundary channel result gamma}\\
\m_{\F^{(q)}}^2&=\[2+\frac{(\s-1)\e}{4}\]+\frac{5+7\s}{96}(N-1)(N+1)^2\a^2+O(\e^{3/2}) \label{Potts boundary channel result mu}
\,,
\ee
where the square-bracketed term is the free-theory solution and $\a^2$ is given in \eqref{Potts alpha}.
The surface critical exponent
\be
\eta_\parallel=2q+2\h\g_{\F^{(q)}}=2q-\frac{(N-1)(6+\s)}{3(3N-7)}\e+O(\e^{3/2})
\,
\ee
agrees with the results in \cite{Diehl1989}.
The solutions for the bulk channel parameters are
\be\label{Potts bulk results}
\g_{\f}&=\frac{(N-1)(N+1)^2}{48}\a^2+O(\e^2)\,,\hspace{23mm}
\g_{\f^2}=-\frac{5(N-1)(N+1)^2}{24}\a^2+O(\e^2)\,, \nn
\l_{\f^2}&=\s+\frac{\s-3}{32}(N-1)(N+1)^2\a^2+O(\e^{3/2})\,, \quad
\l_{\f^3}=-\frac{8+\s}{144}(N-1)(N+1)^2\a^2+O(\e^{3/2})\,.
\ee
For $\D\geq 8$, we have
\be\label{Potts bulk results 2}
\l_{\D}&=\frac{(-1)^{\D/2}(\D-2)\G(\tfrac{\D-4}{2})}{3\times 2^{\D+1}\(1/2\)_{\frac{\D-4}{2}}}(N-1)(N+1)^2\a^2+O(\e^{3/2})
\,.
\ee
The result for the anomalous dimension $\g_\f$ agrees with \cite{Guo:2024bll}.

Let us further discuss some special $N$ limits.
The $N\ra 0$ and $N\ra -1$ limits are associated with percolation and spanning forest at a surface \cite{carton1980surface}.
We find
\be\label{special N}
\h\g_{\F^{(q)}}\big|_{N\ra 0}=-\frac{6+\s}{42}\e+O(\e^{3/2})\,, \qquad
\h\g_{\F^{(q)}}\big|_{N\ra -1}=-\frac{6+\s}{30}\e+O(\e^{3/2})
\,.
\ee
For the $N\ra -1$ limit, the $(N+1)^2$ factor in \eqref{Potts boundary channel result gamma} cancels the $1/(N+1)^2$ pole coming from $\a^2$. 
In contrast to the case without a boundary, the $N\ra\infty$ limit of $\h\g_{\F^{(q)}}$ in the Potts model does not correspond to that in the Yang-Lee case.
This indicates that the different copies $\F^{(q)}_1,\ldots,\F^{(q)}_N$ do not decouple in the $N\ra\infty$ limit.\footnote{See also the bulk discussion in section 5.3.2 of \cite{Wiese:2023vgq}.}
On the other hand, we notice that the coefficient $\l_{\D}$ with $\D\geq 8$ is proportional to the $(-1)^{\D/2}$ part of \eqref{bulk result 2}.
More specifically, the $N\ra\infty$ limit of $\l_{\D}$ ($\D\geq 8$) in the Potts model is exactly the $(-1)^{\D/2}$ part of $\l_{\D}$ in the Yang-Lee model.
This correspondence also holds for higher derivative theories.
Finally, we obtain the one-point function coefficient of $\f^3\equiv\sum_{a,b,c}^{N}T_{abc}\f_a\f_b\f_c$
\be\label{a phi3}
a_{\f^3}=\l_{\f^3}/C_{\f^3}=\frac{8+\s}{108}\a+O(\e)
\,,
\ee
where the bulk OPE coefficient of $\f^3$ is given by \eqref{C}.

\subsection{Higher derivative generalizations}

As in the Yang-Lee case, the matching condition \eqref{bulk boundary 2pt matching} in the Potts model has a higher-derivative generalization,
\be
\lim_{\e\ra 0}\a^{-1}\<\Box^k\f_a(x)\big(\Ps_{\h\D}\big)_b(\h y)\>=\<K_a(x)\big(\Ps_{\h\D}\big)_b(\h y)\>_{\text f}
\,.
\ee
As in the $k=1$ case, the boundary crossing equation does not fix $\a^2$, so we use the result obtained in \cite{Guo:2024bll},
\be\label{alpha-Potts}
\a^2&=-\frac{ (2k)! \, (3k-1)! \, \e }
{ (N+1)^{2}\( (N-2)2^{5-4k}k!+6(N-1)(-1)^k\frac{(1/2)_k^2}{(3k)_k} \) }
+O(\e^{3/2})
\,.
\ee
Let us consider the case of $k=2$.
The Wick contractions imply
\be
F_{2,\text f}(\x)=\frac{2(N-1)(N+1)^2\(1+5\x+10\x^2+10\x^3+(5+3\s_0+2\s_1)\x^4+(1+\s_0)\x^5\)^2}{\x^8 (1+\x)^{10}}
\,.
\ee
The boundary channel expansion gives
\be
\n_{\h\D,\text f}^2&=\begin{cases}
		\; 4(1+\s_0)(N-1)(N+1)^2 &\quad \h\D=8\\
		\; 8(1+\s_0)(1+\s_1)(N-1)(N+1)^2 &\quad \h\D=9\\
		\; \frac{4}{7}(19-9\s_0-7\s_1-35\s_0\s_1)(N-1)(N+1)^2 &\quad \h\D=10\,,
	\end{cases}
\ee
and the squared BOE coefficients with $\h\D\geq 12$ are
\be
\n_{\h\D,\text f}^2=\;&(N-1)(N+1)^2\frac{(\h\D-7)(\h\D-6)(\h\D-4)!(\h\D-1)!}{544320(2\h\D-12)!}\nn
&\times[108(1+(-1)^{\h\D})+(-1)^{\h\D}(\h\D-11)(\h\D-8)(\h\D-3)\h\D(1+\s_0\s_1)]\,.
\ee
These coefficients are $(N-1)(N+1)^2$ times \eqref{k=2 nu} and \eqref{k=2 nu high Delta} in the Yang-Lee case, but the boundary identity contribution is absent here.
Consider the bulk channel expansion
\be
\x^{\D_\f}F_1(\x)=1+\sum_{m=0}^1\l_{\pa^{2m}\f^2} f_\text{bulk}(\D_{\pa^{2m}\f^2},\x)+\sum_{\D=8}^{\infty}\l_{\D} f_\text{bulk}(\D,\x)+O(\e^{3/2})
\,.
\ee
The boundary channel results are
\be
\h\g^{}_{\F^{(q_0)}}&=\frac{N-1}{37N-65}\(21+\frac{55 \sigma _0}{18}+\frac{14 \sigma _1}{9}+\frac{28 \sigma _0 \sigma _1}{3}\)\e^2+O(\e^{3/2})\,, \nn
\h\g^{}_{\F^{(q_1)}}&=-\frac{N-1}{37N-65}\(21+\frac{5 \sigma _0}{2}+4 \sigma _1+42 \sigma _0 \sigma _1\)\e^2+O(\e^{3/2})\,, \label{Potts k=2 boundary results gamma}\\
\m_{\F^{(q_0)}}&=\[6\s_0-4+\frac{37(1-\s_0)}{24}\e\]+\frac{N-1}{37N-65}\(\frac{312701}{1080}-\frac{210449 \sigma _0}{1080}-\frac{23201 \sigma_1}{270}+\frac{36689 \sigma _0 \sigma _1}{270}\)\e \nn
&\hspace{5mm}+O(\e^{3/2})\,,\nn
\m_{\F^{(q_1)}}&=\[7-3\s_1+\frac{7\s_1-11}{8}\e\]+\frac{N-1}{37N-65}\(-\frac{3293}{20}+\frac{24749 \sigma _0}{120}+\frac{4291 \sigma _1}{60}-\frac{14667
   \sigma _0 \sigma _1}{40}\)\e\nn
   &\hspace{5mm}+O(\e^{3/2})
\,. \label{Potts k=2 boundary results mu}
\ee
The bulk channel results are
\be\label{Potts k=2 bulk results}
\g_\f&=-\frac{3(N-1)}{2(37N-65)}\e+O(\e^{3/2})\,, \qquad
\g_{\f^2}=\frac{17(N-1)}{37N-65}\e+O(\e^{3/2})\,, \nn
\g_{\pa^2\f^2}&=-\frac{11(N-1)}{37N-65}\e+O(\e^{3/2})\,,\nn
\l_{\f^2}&=3 \sigma _0+2 \sigma _1-\frac{1}{120(37N-65)} \Big[20685+5081\sigma _0+2834 \sigma _1 +20930\s_0 \sigma
   _1\nn
   &\hspace{37mm}-N \left(20685+5921\sigma _0+3674 \sigma _1+20930 \s_0\sigma
   _1\right)\Big]\e+O(\e^{3/2})\,, \nn
\l_{\pa^2\f^2}&=2 \sigma _0+\frac{2 \sigma _1}{3}-\frac{1}{180
   (37N-65)}\Big[13209+8009\sigma _0+4187 \sigma _1+14315\s_0 \sigma _1\nn
   &\hspace{37mm}-N
   \left(13209+7589\sigma _0+4047 \sigma _1+14315\s_0 \sigma _1\right)\Big]\e+O(\e^{3/2}) \,,\nn
\l_{\f^3}&=-\frac{N-1}{84(37N-65)} (672+65\sigma _0+44 \sigma _1+504 \s_0\sigma _1)\e+O(\e^{3/2})\,,\nn
\l_{\pa^2\f^3}&=\frac{N-11}{882(37N-65)} (1092+25\sigma _0 +25 \sigma _1+1008 \s_0\sigma _1)\e+O(\e^{3/2})
\,.
\ee
For $\D\geq 16$, we find
\be\label{Potts k=2 bulk results 2}
\l_{\D}=&-\frac{\sqrt{\pi } (\Delta -6)^2 (\Delta -4) (\Delta -2)\Gamma (\frac{\Delta
   }{2}-5)
}{27\times 2^{\Delta+5} \Gamma (\frac{\Delta -5}{2})}\frac{N-1}{37N-65}\nn
   &\times(-1)^{\Delta/2} (\D(\D-12)(\D(\D-12)+188) (\s_0\s_1  +1)+576 (8 \s_0\s_1 +11))\e+O(\e^{3/2})
   \,,
\ee
where $\D$ is an integer.
The bulk anomalous dimensions agree with \cite{Guo:2024bll}.
As mentioned below \eqref{special N}, the $N\ra\infty$ limit of $\l_{\D}$ with $\D\geq 16$ corresponds to the $(-1)^{\D/2}$ part of \eqref{YL k=2 bulk results 2}.
The one-point function coefficient of $\f^3$ is
\be\label{a phi3 k=2}
a_{\f^3}=\pm\frac{672+65\s_0+44\s_1+504\s_0\s_1}{42(N+1)\sqrt{35(65-37N)}}\e^{1/2}+O(\e)
\,.
\ee
We also obtain the $k=3$ results, which are given in Appendix \ref{Results for the k=3 case}.

\section{Towards $\f^{2n+1}$ theories}
\label{Towards phi2n+1 theories}

In this section, we attempt to generalize the procedure to the $\f^{2n+1}$ theories with $n>1$.
Recall that there are two steps: first using the multiplet recombination and then using boundary crossing symmetry.
For general $(k,n)$,\footnote{To avoid mixing with derivative bulk interactions, $k$ and $2n-1$ should be coprime \cite{Guo:2024bll}.}
it is straightforward to generalize the first step, i.e., to obtain the BOE coefficients that generalizes \eqref{mu result}.
This discussion corresponds to Sec. \ref{General n>1 theories}.
The resummation or differential equation is more complicated.
To proceed, we consider the $(k,n)=(1,2)$ case as an example.
Unfortunately, the generalization of the second step does not lead to strong enough constraints.
As we will discuss in Sec. \ref{phi5 theory}, the crossing constraints are insufficient to determine the anomalous dimension of the boundary fundamental operator.
A similar situation also exists in the $\f^{2n+1}$ theory without a boundary.
In that case \cite{Guo:2024bll}, 
the methods based on the multiplet recombination and Lorentzian inversion formula cannot determine the anomalous dimension of $\f$.

\subsection{General $n>1$ theories}
\label{General n>1 theories}

Let us consider the higher derivative and multicritical generalization of \eqref{bulk boundary matching},
\be
\lim_{\e\ra 0}\a^{-1}\<\Box^k \f(x)\Ps_{\h\D}(\h y)\>=\<\f^{2n}(x)\Ps_{\h\D}(\h y)\>_{\text f}
\,.
\ee
Here the upper critical dimension is $d_{\text u}=2k\frac{2n+1}{2n-1}$.
The LHS of the matching condition above is again given by that of \eqref{boxk 2pt}.
It can be shown that $\g_\f$ is of order $\a^2$ for general $(k,n)$ \cite{Guo:2024bll}, so only the $l=0$ term in \eqref{boxk 2pt} survives the $\e\ra 0$ limit.
We have
\be\label{general BOE coefficient}
\m_{\h\D}=\frac{\n_{\h\D,\text f}}{4^{k} (\frac{2k}{2n-1}-\h\D)_{2k}}\a+O(\e)
\,.
\ee
The one-point function coefficient $a_\f$ can be obtained by considering the $\h\D=0$ case.
The free BOE coefficient $\n_{j,\text f}$ on the RHS is encoded in
\be
\<\f^{2n}(x)\f^{2n}(y)\>_{\text f}=\sum_{i=0}^n (2n-2i)!\[\binom{2n}{2i}(2i-1)!!\]^2 \Big(\<\f(x)\f(y)\>_{\text f}\Big)^{2n-2i}
\Big(\<\f^2(x)\>_{\text f}\<\f^2(y)\>_{\text f}\Big)^i
\,.
\ee
One can in principle obtain any $\n_{\h\D,\text f}$ from the boundary channel expansion of this correlator.
For $\h\D=0$, we have $\n_{\h\D,\text f}=(2n-1)!! \sum_{j=0}^{k-1}B_j$, where $B_j$ is given by \eqref{B}.
So the one-point function coefficient of $\f$ is
\be\label{general a phi}
a_\f\equiv\m_0=\frac{(2n-1)!! \sum_{j=0}^{k-1}B_j}{4^{k} (\frac{2k}{2n-1})_{2k}}\a+O(\e)
\,,
\ee
which generalizes the formula \eqref{a}.\footnote{
For $n=1$, the formula for $\a$ can be found in \cite{Guo:2024bll}:
\be
	\a^{2}&=-\frac{ (2k)! \, (3k-1)! \, \e }
	{ 2^{5-4k}k!+6(-1)^{k}\frac{ \(1/2\)_{k}^{2} }{ \(3k\)_{k} } }
	+O(\e^{3/2})
	\,.
\ee}
To be more concrete, below we consider the $(k,n)=(1,2)$ case, i.e., the canonical $\f^5$ theory with a boundary.
As opposed to the $n=1$ case, we are not able to fix $\h\g_{\F^{(q)}}^{}$ by boundary crossing symmetry.

\subsection{$\f^5$ theory}
\label{phi5 theory}

A simple indication for the technical difficulty in the $\f^5$ theory is a fractional upper critical dimension
\be 
d_{\text u}=\frac{10}{3}\,.
\ee
Accordingly, at order $\e^0$, the scaling dimensions $\h\D$ are fractions instead of integers.
It is more challenging to resum the leading corrections to the boundary channel expansion, 
so we turn to the differential equation approach.
The corresponding matching condition is
\be\label{bulk bulk matching phi5}
\lim_{\e\ra 0}\a^{-2}\<\Box\f(x)\Box\f(y)\>=\<\f^{4}(x)\f^{4}(y)\>_{\text f}
\,.
\ee
The free correlator takes the form
\be\label{diff-eq-phi5}
\<\f^{4}(x)\f^{4}(y)\>_{\text f}=\frac{F_{4,\text f}(\x)}{(4x_\perp y_\perp)^{4\D_{\f,\text f}}}
\,,
\ee
where $F_{4,\text f}(\x)$ is given by
\be
F_{4,\text f}(\x)=9+72\(\frac{1}{\x^{\frac{d-2}{2}}}+\frac{1}{(1+\x)^{\frac{d-2}{2}}}\)^2+24\(\frac{1}{\x^{\frac{d-2}{2}}}+\frac{1}{(1+\x)^{\frac{d-2}{2}}}\)^4
\,.
\ee
It is difficult to fully solve the fourth-order differential equation \eqref{bulk bulk matching phi5}. 
Note that the fourth order differential operator in \eqref{D} can be factorized as
\be
\cal D=16\tilde{\cal D}_2(\D_\f, {y_\perp}/{x_\perp})\tilde{\cal D}_1(\D_\f, {x_\perp}/{y_\perp}), 
\ee
where the second order differential operators are
\be
\tilde{\cal D}_1(\D_\f,z)&=\x(1+\x)\pa_\x^2+\[\frac{d}{2}(1+2\x)+\(\D_\f-\frac{d-2}{2}\)\(1+2\x-z\)\]\pa_\x+\D_\f(\D_\f+1)\,,\nn
\tilde{\cal D}_2(\D_\f,z)&=\tilde{\cal D}_1(\D_\f, z)+\frac{2(\D_\f+1)}{z}\pa_z+\frac{1}{z^2}\pa^2+((1+2\x)z-1)\pa_\x\pa_z
\,.
\ee
Therefore, we can partially solve the fourth order differential equation by considering the second-order differential equation associated with only $\Box_y$. 
The action of $\Box_x$ on the bulk two-point function $\<\f(x)\f(y)\>$ gives
\be
\<\Box\f(x)\f(y)\>=\frac{G(\x, {x_\perp}/{y_\perp})}{4^{\D_\f}x_\perp^{\smash{\D_\f+2}}y_\perp^{\smash{\D_\f}}}
\,,
\ee
where we have $G(\x, {x_\perp}/{y_\perp})=\tilde{\cal D}(\D_\f, {x_\perp}/{y_\perp})F_1(\xi)$. 
Then \eqref{bulk bulk matching phi5} reduces to a second-order differential equation
\be
\tilde{\cal D}_2(\D_\f, {y_\perp}/{x_\perp})\, G(\x, {x_\perp}/{y_\perp})=\a^2 F_{4,\text f}(\x)+O(\e^{3/2})
\,.
\ee
The solution reads
\be
G=\;&\frac{9 \alpha ^2}{10 \xi ^{5/3}
   (1+\xi )^{5/3}} \bigg[320 \xi ^{2/3}+16 (1+\xi )^{2/3}-720 \xi ^{4/3}
   (1+\xi)^{1/3} \left(2+(1+\xi)^{1/3}\right)\nn
   &-720 \xi ^{7/3} (1+\xi)^{1/3}
   \left(2+(1+\xi)^{1/3}\right)-\xi ^2 \left(720-456 (1+\xi )^{2/3}\right)-8 \xi 
   \left(90-19 (1+\xi )^{2/3}\right)\nn
   &+3 \xi ^{8/3} \left(152+240 (1+\xi)^{1/3}+3
   (1+\xi )^{2/3}\right)+\xi ^{5/3} \left(760+720 (1+\xi)^{1/3}+9 (1+\xi
   )^{2/3}\right)\nn
   &+360 \xi ^{5/3} (1+\xi ) \left(\xi ^{2/3}-(1+\xi )^{2/3}\right) \,
   _2F_1\left(\frac{1}{3},\frac{2}{3};\frac{5}{3};-\xi \right)\bigg]+\frac{c_1}{\x^{2/3}}+\frac{c_2}{(1+\x)^{2/3}}+O(\e^{3/2})
\,,
\ee
where the ratio ${x_\perp}/{y_\perp}$ may appear at higher orders in $\e$. 
Since $\Box\f_\text{f}=0$ (ignoring contact terms), the function $G$ vanishes in the $\e\ra 0$ limit, and $(c_1, c_2)$ are of positive order in $\e$. 
Below, we focus on the Neumann case as an example.
The boundary channel expansion implies
\be
c_1&=-\h\g_{\F^{(0)}}+\frac{27\a^2}{80}\(44+\frac{15\times 2^{2/3} \Gamma
   \left(-\frac{1}{6}\right) \Gamma
   \left(\frac{5}{3}\right)}{\sqrt{\p }}\)+O(\e^{3/2})\,,\nn
c_2&=-\h\g_{\F^{(0)}}+\frac{27\a^2}{80}\(44-\frac{15\times 2^{2/3} \Gamma
   \left(-\frac{1}{6}\right) \Gamma
   \left(\frac{5}{3}\right)}{\sqrt{\p }}\)+O(\e^{3/2})
\,.
\ee
The bulk channel expansion for $G(\x)$ reads\footnote{For the infinite sum on the RHS, the subscripts correspond to the schematic forms of the bulk primaries.
Since the scaling dimensions are not integers, we write these schematic expressions for clarity, as opposed to the $n=1$ case.}
\be\label{phi5 crossing}
G(\x, {x_\perp}/{y_\perp})
=\;&\tilde{\cal D}_1(\D_\f, {x_\perp}/{y_\perp})\bigg[\x^{-\D_\f}\Big(1+\l_\f f_\text{bulk}(\D_\f,\x)+\l_{\f^2} f_\text{bulk}(\D_{\f^2},\x) +\l_{\f^3} f_\text{bulk}(\D_{\f^3},\x)\nn
&\hspace{7mm}+\l_{\f^{5}} f_\text{bulk}(\D_{\f^5},\x)+\sum_{p=3}^5\sum_{m=1}^{\infty}\l_{\pa^{2m}\f^p} f_\text{bulk}(\D_{\pa^{2m}\f^p},\x)+O(\e^{3/2})\Big)\bigg]
\,.
\ee
We would like to briefly comment on this expansion.
The bulk operator $\f^4$ is a descendant, and its contribution is absorbed into the $\f$ block.
In contrast to the $\f^3$ theory case, the primary $\f^4_{\text f}$ does not contribute in the free theory.
This means that $\l_\f f_\text{bulk}(\D_\f,\x)$ vanishes in the $\e\ra 0$ limit.
Since $f_\text{bulk}(\D_\f,\x)$ diverges as $1/\g_\f$, the coefficient $\l_\f$ should be of higher order than $\g_\f$.
We find the relation
\be\label{phi5 solution}
\l_{\f^3}&=-3\h\g_{\F^{(0)}}+\frac{27\a^2}{80}\(\frac{45\times 2^{2/3} \Gamma
   \left(-\frac{1}{6}\right) \Gamma
   \left(\frac{5}{3}\right)}{\sqrt{\p }}-148\)+O(\e^{3/2})\,, \nn
\l_{\f^5}&=-\frac{3\h\g_{\F^{(0)}}}{5}+\frac{27\a^2}{800}\(593-\frac{90\times 2^{2/3} \Gamma
   \left(-\frac{1}{6}\right) \Gamma
   \left(\frac{5}{3}\right)}{\sqrt{\p}}\)+O(\e^{3/2})
\,.
\ee
However, we are not able to determine $\h\g_{\F^{(0)}}$ by these relations because we do not know the bulk one-point functions coefficients in $\l_{\f^3}$ and $\l_{\f^5}$.
Let us further explore the constraints from the full crossing symmetry.
In other words, we consider the crossing equation for the original two-point function $\<\f(x)\f(y)\>=\frac {F_1(\x)} {(4x_\perp y_\perp)^{\D_\f}}$, instead of $\<\Box\f(x)\f(y)\>$.
Substituting \eqref{phi5 solution} into the bulk channel expansion
\be\label{full crossing}
\x^{\D_\f}F_1(\x)
=\;&1+\l_\f f_\text{bulk}(\D_\f,\x)+\l_{\f^2} f_\text{bulk}(\D_{\f^2},\x) +\l_{\f^3} f_\text{bulk}(\D_{\f^3},\x)\nn
&+\l_{\f^{5}} f_\text{bulk}(\D_{\f^5},\x)+\sum_{p=3}^5\sum_{m=1}^{\infty}\l_{\pa^{2m}\f^p} f_\text{bulk}(\D_{\pa^{2m}\f^p},\x)+O(\e^{3/2})
\,,
\ee
the remaining unknown parameters include $\h\g_{\F^{(q)}}$ and the leading corrections to $(\m_{\F^{(0)}}^2, \l_\f^2)$, which will be denoted as $(\m_{\F^{(0)}}^2)^{(1)}$ and $\l_\f^{(1)}$.
In the bulk channel expansion \eqref{full crossing}, the terms associated with the unknown parameters are 
\be
&-\l_\f^{(1)}(1+\x)^{2/3}\,\e+
\frac{1}{2}(\m_{\F^{(0)}}^2)^{(1)}\[(1+\x)^{2/3}+(2+\x+1/\x)^{2/3}\]\e\nn
&+\frac{1}{6}\h\g_{\F^{(0)}}\[(\sqrt 3\p+9\log 3-9)(1+\x)^{2/3}-(\sqrt 3\p-9\log 3)(2+\x+1/\x)^{2/3}\]\e
\,,
\ee
where we have dropped an overall factor $\x^{2/3}(1+\x)^{-4/3}$.
One can see that these three terms are not linearly independent. 
As the $(\m_{\F^{(0)}}^2)^{(1)}$ and $\h\g_{\F^{(0)}}$ terms in $F_1(\x)$ are annihilated by the differential operator \eqref{D}, 
they are not constrained by the differential equation \eqref{bulk bulk matching phi5}. 
Therefore, we can shift the anomalous dimension $\h\g_{\F^{(0)}}$ arbitrarily, and it remains a free parameter.

\section{Discussion}

In this paper, we considered the $\f^3$ theory with a boundary and their higher-derivative generalizations.
We used the multiplet recombination in conjunction with the boundary crossing symmetry, and we determined the anomalous dimensions of the boundary fundamental operators.
For the $\f^{2n+1}$ theories with $n>1$, we only determined some leading BOE coefficients.

For the $\f^3$ theory, both the single field and $S_{N+1}$-symmetric multi-field cases are studied.
In the CFT without a boundary, the large $N$ limit of the multi-field case corresponds to $N+1$ decoupled theories with a single field \cite{Wiese:2023vgq}.
This is no longer the case in the presence of a boundary.
In some sense, the boundary acts as a nonlocal probe, allowing us to distinguish CFTs that local probes alone cannot tell apart.
 
In $\f^{2n}$ theories with a boundary, the anomalous dimensions of the boundary composite operators are found in \cite{Nishioka:2022odm,Guo:2025edk}.
This is achieved using the analyticity of bulk-boundary-boundary three-point functions $\<\f\,\F^p\,\F^{p+1}\>$.
The direct generalization to the $\f^3$ theory fails for the following reasons.
In the conformal block expansion, only $\F$ contributes in the free theory.
There are infinitely many new contributions in the interacting theory, but we do not know the new contributions associated with the $\f^3$ interaction.
The correlator $\<\f\,\F^p\,\F^{p+2}\>$ does not help either because it vanishes in the free theory and the anomalous dimensions do not appear in the leading correction.
Perhaps one needs to examine bulk-bulk-boundary three-point functions, but their conformal block expansions are more complicated.

We did not fix the anomalous dimension of the boundary fundamental operator in the $\f^{2n+1}$ theory for $n>1$.
Perhaps it could be determined using input from the Lagrangian.
More broadly, it would be interesting to incorporate Lagrangians in the bootstrap studies.
This might help sharpen the bootstrap predictions.

As mentioned in the introduction, the motivation for studying the single field $\f^3$ CFT was its connection to branched polymers.
The corresponding boundary CFT naturally describes branched polymers at a surface.
But the single field $\f^3$ theory is also related to the Yang-Lee theory.
Here, we briefly discuss this relation in the presence of a boundary.
In the thermodynamic limit, the Yang-Lee zeros are associated with a line of singularities in the free energy as a function of the bulk external magnetic field.
Near the edge of the line, the scaling dimension of $\f$ controls the singular behavior of the free energy, i.e., the density of Yang-Lee zeros.
If another external magnetic field is imposed on the boundary, then the free energy should exhibit certain singular behavior controlled by the scaling dimension of $\F$.
In other words, the scaling dimension of $\F$ should characterize the behavior of the density of zeros as we vary the boundary external magnetic field.

The coefficients in the bulk channel expansion are products of bulk OPE and one-point function coefficients.
These products are not necessarily positive, so the numerical bootstrap that relies on positivity conditions would not work.
It is interesting to explore the bootstrap methods that apply to this scenario \cite{Gliozzi:2015qsa,Padayasi:2021sik,Hu:2025yrs}.
Moreover, it would also be useful to develop novel bootstrap methods for CFTs without a boundary \cite{Gliozzi:2013ysa,Gliozzi:2014jsa,El-Showk:2016mxr,Esterlis:2016psv,Li:2017agi,Li:2017ukc,Li:2021uki,Kantor:2021kbx,Kantor:2021jpz,Afkhami-Jeddi:2021iuw,Laio:2022ayq,Li:2023tic}.

Interface CFTs can be represented by boundary CFTs using the folding trick \cite{WONG1994403}.
See e.g. \cite{Gaiotto:2012np,Gliozzi:2015qsa,Dey:2020jlc,Bartlett-Tisdall:2025kcx} for more recent studies.
It would be interesting to generalize the procedure to interface CFTs.
Of particular interests are the renormalization group domain walls, which can be used to study the relation between UV and IR CFTs.

\section*{Acknowledgments}
This work was supported by the Natural Science Foundation of China (Grant No. 12522504 and No. 12205386).

\appendix
\section{Free $\Box^k$ boundary conformal field theory}

In this appendix, we review the free $\Box^k$ BCFT \cite{Chalabi:2022qit}.
In this theory, there are $k$ boundary conditions, which are labeled using $i=0,1,\ldots,k-1$.
We introduce 
\be
\s_i=\begin{cases}
	\; +1 \quad & \text{(N)} \\
	\; -1  & \text{(D)}\,,
\end{cases}
\ee
where N (D) represents the boundary condition of the Neumann (Dirichlet) type.
Each boundary condition kills one of the two operators $\{\F^{(i)},\F^{(2k-1-i)}\}$.
The remaining operators form the set of boundary fundamental primaries.
The scaling dimension of a boundary fundamental primary is
\be
\h\D_i=\frac {d-2k} {2}+q_i
\,,
\ee
where the number of derivatives $q_i$ is related to the boundary condition by
\be
q_i=i\frac{1+\s_i}{2}+(2k-1-i)\frac{1-\s_i}{2}
\,.
\ee
The bulk two-point function takes the form
\be
\<\phi(x)\phi(y)\>=\frac{F_1(\x)}{\(4\xn\yn\)^{\D_\phi}}
\,,
\ee
where the function $F_1(\x)$ is
\be\label{free F}
F_1(\x)=\frac{1}{\x^{\frac{d}{2}-k}}+\sum_{j=0}^{k-1}\frac{B_j}{(\x+1)^{\frac{d}{2}-k+j}}
\,,
\ee
with the coefficients \cite{Guo:2025edk}
\be\label{B}
B_j=\;\(\frac{d-2k}{2}\)_j\,\sum _{i=0}^{j}\frac{(2i-2k+1)(1-k)_j}{i!(j-i)!(i-2k+1)_{j+1}}\,\s_i
\,.
\ee

\section{Results for the $k=3$ case}
\label{Results for the k=3 case}

In the Yang-Lee case, the boundary channel results are
\be
\a^2&=-1135411200\e+O(\e^{3/2})\,,\\
\h\g^{}_{\F^{(q_0)}}&=-\(\frac{469}{144}+\frac{763}{1560}\s_0+\frac{11}{30}\s_1+\frac{539}{4680}\s_2+\frac{271}{80}\s_0\s_1+\frac{7}{5}\s_0\s_2+\frac{77}{780}\s_0\s_1\s_2\)\e+O(\e^{3/2})\,, \nn
\h\g^{}_{\F^{(q_1)}}&=\(\frac{1625}{144}+\frac{245}{312}\s_0+\frac{320}{351}\s_1+\frac{245}{936}\s_2+\frac{1355}{144}\s_0\s_1+\frac{175}{36}\s_1\s_2+\frac{385}{1404}\s_0\s_1\s_2\)\e+O(\e^{3/2})\,, \nn
\h\g^{}_{\F^{(q_2)}}&=-\(\frac{295}{36}+\frac{5}{13}\s_0+\frac{5}{13}\s_1+\frac{140}{117}\s_2+15\s_0\s_2+\frac{75}{4}\s_1\s_2+\frac{55}{52}\s_0\s_1\s_2\)\e+O(\e^{3/2})\,, \label{YL k=3 boundary results gamma}\\
\m_{\F^{(q_0)}}&=\[22-20\s_0+\frac{1627(\s_0-1)}{240}\e\]+\bigg(\frac{124075957}{453600}-\frac{108645431 \sigma _0}{368550}-\frac{2231395 \sigma
   _1}{12096}-\frac{131197663 \sigma
   _2}{3369600} \nn
   &\hspace{5mm}+\frac{631255267 \sigma _0 \sigma _1}{3931200}+\frac{19564327 \sigma _0 \sigma _2}{673920}+\frac{70614421 \sigma _1
   \sigma _2}{1123200}-\frac{71274061 \sigma _0 \sigma _1 \sigma _2}{1123200}\bigg)\e+O(\e^{3/2})\,, \nn
\m_{\F^{(q_1)}}&=\[-60+66\s_1-\frac{279\s_1-271}{16}\e\]+\bigg(\frac{6152519}{8064}-\frac{12013613 \sigma _0}{24192}-\frac{202101941 \sigma
   _1}{314496}-\frac{6502039 \sigma
   _2}{22464}\nn
   &\hspace{5mm}+\frac{186441097 \sigma _0 \sigma _1}{314496}-\frac{141809 \sigma _0 \sigma _2}{11232}+\frac{7660697 \sigma _1 \sigma
   _2}{22464}+\frac{55957 \sigma _0 \sigma _1 \sigma _2}{3744}\bigg)\e+O(\e^{3/2})\,,\nn
\m_{\F^{(q_2)}}&=\[35-21\s_2+\frac{30\s_2-43}{6}\e\]+\bigg(-\frac{5486459}{11232}+\frac{5684665 \sigma _0}{7488}+\frac{2356585 \sigma
   _1}{2496}+\frac{3782581 \sigma
   _2}{11232}\nn
   &\hspace{5mm}+\frac{219367 \sigma _0 \sigma _1}{3744}-\frac{2512933 \sigma _0 \sigma _2}{2496}-\frac{9412493 \sigma _1 \sigma
   _2}{7488}-\frac{92819 \sigma _0 \sigma _1 \sigma _2}{1248}\bigg)\e+O(\e^{3/2})
\,. \label{YL k=3 boundary results mu}
\ee
The parameters in the bulk channel are determined to be
\be\label{YL k=3 bulk results}
\g_\f&=-\frac{5}{36}\e+O(\e^{3/2})\,, \quad
\g_{\pa^2\f^2}=-\frac{65}{36}\e+O(\e^{3/2})\,, \quad
\g_{\pa^4\f^2}=\frac{17}{18}\e+O(\e^{3/2})\,, \nn
\l_\f&=\(\frac{\sigma _0}{288}+\frac{\sigma _1}{160}+\frac{7 \sigma _2}{2160}\)\e+
\bigg(\frac{23726951}{344881152}+\frac{7638173 \sigma_0}{1132185600}+\frac{4230101 \sigma _1}{377395200}+\frac{7457 \sigma _2}{673920}\nn
&\hspace{5mm}+\frac{350503 \sigma _0 \sigma _1}{35481600}+\frac{8409751 \sigma _0 \sigma _2}{136857600}+\frac{24407 \sigma _1 \sigma_2}{202752}+\frac{77221 \sigma _0 \sigma _1 \sigma_2}{13478400}\bigg)\e^2+O(\e^{5/2})\,, \nn
\l_{\pa^2\f^2}&=\frac{25 \sigma _0}{2}+\frac{33 \sigma _1}{2}+7 \sigma _2+\bigg(\frac{78888805}{798336}+\frac{4172771 \sigma_0}{943488}+\frac{13985609 \sigma _1}{943488}+\frac{364421 \sigma _2}{16848}\nn
&\hspace{5mm}-\frac{3375733 \sigma _0 \sigma _1}{88704}+\frac{14631413 \sigma _0 \sigma _2}{114048}+\frac{10069513 \sigma _1 \sigma_2}{38016}+\frac{387997 \sigma _0 \sigma _1 \sigma_2}{33696}\bigg)\e+O(\e^{3/2})\,, \nn
\l_{\pa^4\f^2}&=\frac{7 \sigma _0}{3}+\frac{7 \sigma _1}{5}+\frac{7 \sigma _2}{15}+\bigg(-\frac{13260491}{1710720}-\frac{6827689 \sigma_0}{1263600}-\frac{5489773 \sigma _1}{1263600}-\frac{1380721 \sigma _2}{2527200}\nn
&\hspace{5mm}-\frac{16240121 \sigma _0 \sigma _1}{950400}+\frac{21627971 \sigma _0 \sigma _2}{4276800}+\frac{791149 \sigma _1 \sigma_2}{57024}+\frac{32683 \sigma _0 \sigma _1 \sigma_2}{78975}\bigg)\e+O(\e^{3/2})\,, \nn
\l_{\f^3}&=\bigg(\frac{25973}{14256}+\frac{35 \sigma _0}{286}+\frac{95 \sigma_1}{858}+\frac{1127 \sigma_2}{23166}+\frac{271 \sigma _0 \sigma _1}{176}+\frac{7 \sigma _0 \sigma _2}{11}\nn
&\hspace{8mm}+\frac{35 \sigma _1 \sigma _2}{44}+\frac{7\sigma _0 \sigma _1 \sigma_2}{156}\bigg)\e+O(\e^{3/2})\,, \nn
\l_{\pa^2\f^3}&=-\(\frac{26}{55}+\frac{301 \sigma _0}{41184}+\frac{469 \sigma _1}{61776}+\frac{931 \sigma _2}{205920}+\frac{1439
   \sigma _0 \sigma _1}{3960}+\frac{5579 \sigma _0 \sigma
   _2}{31680}+\frac{595 \sigma _1 \sigma _2}{2112}+\frac{119 \sigma _0 \sigma _1
   \sigma _2}{28080}\)\e\nn
   &\hspace{7mm}+O(\e^{3/2})\,,
\nn
\l_{\pa^4\f^3}&=\bigg(\frac{429149}{3397680}+\frac{19733 \sigma _0}{88339680}+\frac{31703 \sigma
   _1}{132509520}+\frac{17983 \sigma
   _2}{88339680}+\frac{164129 \sigma _0 \sigma _1}{1698840}\nn
   &\hspace{8mm}+\frac{72233 \sigma _0 \sigma _2}{1510080}+\frac{345121 \sigma _1
   \sigma _2}{4530240}+\frac{2261 \sigma _0 \sigma _1 \sigma _2}{12046320}\bigg)\e+O(\e^{3/2})
\,,
\ee
and for $\D\geq 24$,
\be\label{YL lambda k=3 result}
\l_\D&=-\frac{\sqrt{\pi } (\Delta -8) \Gamma \left(\frac{\Delta
   }{2}\right)}{2^{\Delta }(\Delta -16) (\Delta -14) (\Delta -12) \Gamma \left(\frac{\Delta
   -9}{2}\right)}\Bigg[\frac{(1-\D/2)_8}{495} \left(\frac{\sigma _0 \sigma _1}{14}+\frac{\sigma _2 \sigma
   _1}{15}+\frac{\sigma _0 \sigma _2}{27}+\frac{115}{1134}\right)\nn
  &\quad +\frac{(-1)^{\Delta/2}}{5054400}\Big(\frac{1}{6} (4054302720-2780559360 \Delta +696327936 \Delta ^2-96690816 \Delta
   ^3+8488048 \Delta ^4\nn
   &\quad-501120 \Delta ^5+19864 \Delta ^6-504 \Delta ^7+7 \Delta
   ^8)+\frac{1}{2} (1121402880-768890880 \Delta +171134208 \Delta ^2\nn
   &\quad-21080448
   \Delta ^3+1636624 \Delta ^4-86400 \Delta ^5+3112 \Delta ^6-72 \Delta ^7+\Delta
   ^8) \sigma _0 \sigma _1+\frac{1}{3} (704471040\nn
   &\quad-533191680 \Delta +132223488
   \Delta ^2-17512128 \Delta ^3+1440304 \Delta ^4-79920 \Delta ^5+2992 \Delta ^6-72
   \Delta ^7\nn
   &\quad+\Delta ^8) \sigma _0 \sigma _2+(301916160-256711680 \Delta
   +74562048 \Delta ^2-11412288 \Delta ^3+1076464 \Delta ^4\nn
   &\quad-66960 \Delta ^5+2752 \Delta
   ^6-72 \Delta ^7+\Delta ^8) \sigma _1 \sigma _2\Big)\Bigg]\e+O(\e^{3/2})
\,.
\ee
The anomalous dimensions $\g_\f$, $\g_{\pa^2\f^2}$, and $\g_{\pa^4\f^2}$ agree with the results derived before \cite{Guo:2024bll}.
We also find the one-point function coefficient 
\be\label{a phi3 YL k=3}
a_{\f^3}=\l_{\f^3}/C_{\f^3}&=\pm\frac{30}{\sqrt{-77}}\bigg(\frac{25973}{14256}+\frac{35 \sigma _0}{286}+\frac{95 \sigma_1}{858}+\frac{1127 \sigma_2}{23166}+\frac{271 \sigma _0 \sigma _1}{176}+\frac{7 \sigma _0 \sigma _2}{11}\nn
&\hspace{8mm}+\frac{35 \sigma _1 \sigma _2}{44}+\frac{7\sigma _0 \sigma _1 \sigma_2}{156}\bigg)\e^{1/2}+O(\e)
\,.
\ee

For the $k=3$ Potts model, we find
\be
\h\g_{\F^{(q_0)}}^{}&=-\frac{N-1}{6N-17}\(\frac{121}{6}+\frac{763 \sigma _0}{260}+\frac{11 \sigma _1}{5}+\frac{539 \sigma _2}{780}+\frac{429 \sigma _0
   \sigma _1}{20}+\frac{539 \sigma _0 \sigma
   _2}{60}+\frac{77\sigma _0 \sigma _1 \sigma _2}{130}\)\e+O(\e^{3/2})\,, \nn
\h\g_{\F^{(q_1)}}^{}&=\frac{N-1}{6N-17}\(\frac{220}{3}+\frac{245 \sigma _0}{52}+\frac{640 \sigma _1}{117}+\frac{245 \sigma _2}{156}+\frac{715 \sigma _0
   \sigma _1}{12}+\frac{385 \sigma _1 \sigma
   _2}{12}+\frac{385}{234} \sigma _0 \sigma _1 \sigma _2\)\e+O(\e^{3/2})\,, \nn
\h\g_{\F^{(q_2)}}^{}&=-\frac{N-1}{6N-17}\(55+\frac{30 \sigma _0}{13}+\frac{30 \sigma _1}{13}+\frac{280 \sigma _2}{39}+\frac{385
   \sigma _0 \sigma _2}{4}+\frac{495 \sigma _1 \sigma _2}{4}+\frac{165}{26} \sigma _0
   \sigma _1 \sigma _2\)\e+O(\e^{3/2})\,, \label{Potts k=3 boundary results gamma}\\
\m_{\F^{(q_0)}}&=\[22-20\s_0+\frac{1627(\s_0-1)}{240}\e\]+\frac{N-1}{6N-17}\bigg(\frac{2501994653}{1965600}-\frac{2761969213 \sigma _0}{1965600}-\frac{119045847 \sigma
   _1}{72800} \nn
   &\hspace{5mm}+\frac{12958495 \sigma _0 \sigma _1}{8736}-\frac{99825103 \sigma
   _2}{140400}+\frac{45430693 \sigma _0 \sigma _2}{70200}-\frac{235257 \sigma _1 \sigma
   _2}{5200}+\frac{650801 \sigma _0 \sigma _1 \sigma _2}{15600}\bigg)\e+O(\e^{3/2})\,, \nn
\m_{\F^{(q_1)}}&=\[66\s_1-60-\frac{279\s_1-271}{16}\e\]+\frac{N-1}{6N-17}\bigg(\frac{854233}{168}-\frac{1625221 \sigma _0}{504}-\frac{3519137 \sigma
   _1}{819}-\frac{919805 \sigma
   _2}{468} \nn
   &\hspace{5mm}+\frac{50300893 \sigma _0 \sigma _1}{13104}-\frac{141809 \sigma _0 \sigma _2}{1872}+\frac{4322431 \sigma _1 \sigma
   _2}{1872}+\frac{55957}{624} \sigma _0 \sigma _1 \sigma _2\bigg)\e+O(\e^{3/2})\,,\nn
\m_{\F^{(q_2)}}&=\[35-21\s_2+\frac{30\s_2-43}{6}\e\]+\frac{N-1}{6N-17}\bigg(-\frac{6209909}{1872}+\frac{6063875 \sigma _0}{1248}+\frac{2584111 \sigma
   _1}{416}+\frac{479275 \sigma
   _2}{208}\nn
&\hspace{5mm}+\frac{219367 \sigma _0 \sigma _1}{624}-\frac{2685183 \sigma _0 \sigma _2}{416}-\frac{10342643 \sigma _1 \sigma
   _2}{1248}-\frac{92819}{208} \sigma _0 \sigma _1 \sigma _2\bigg)\e+O(\e^{3/2})
\,, \label{Potts k=3 boundary results mu}
\ee
where we have used \eqref{alpha-Potts} and terms in the square brackets are the free-theory solutions.
The parameters in the bulk channel are
\be
\g_\f&=-\frac{5(N-1)}{6(6N-17)}\e+O(\e^{3/2})\,,\qquad
\g_{\f^2}=\frac{23(N-1)}{6(6N-17)}\e+O(\e^{3/2})\,, \qquad \nn
\g_{\pa^2\f^2}&=-\frac{65(N-1)}{6(6N-17)}\e+O(\e^{3/2})\,, \qquad
\g_{\pa^4\f^2}=\frac{17(N-1)}{3(6N-17)}\e+O(\e^{3/2})\,,\nn
\l_{\f^2}&=\frac{15 \sigma _0}{2}+\frac{27 \sigma _1}{2}+7 \sigma _2-\frac{1}{786240 (6N-17)}\bigg[739695538+112355999 \sigma _0+139138865 \sigma _1\nn
&\hspace{5mm}+30405250 \sigma _2+52380910 \sigma
   _0 \sigma _1+726302996 \sigma _0 \sigma _2+1413725732 \sigma _1 \sigma _2-3726386
   \sigma _0 \sigma _1 \sigma _2\nn
&\hspace{5mm}+N (-741046888-117851489 \sigma
   _0-154003715 \sigma _1-41125960 \sigma _2-53732260 \sigma _0 \sigma _1\nn
&\hspace{14mm}-724951646
   \sigma _0 \sigma _2-1412374382 \sigma _1 \sigma _2+2375036 \sigma _0 \sigma _1
   \sigma _2)\bigg]\e+O(\e^{3/2})
\,,\nn
\l_{\pa^2\f^2}&=\frac{25 \sigma _0}{2}+\frac{33 \sigma _1}{2}+7 \sigma _2+\frac{1}{157248 (6N-17)}\bigg[-111780240-3992591 \sigma _0-13228853 \sigma _1\nn
&\hspace{5mm}-20047216 \sigma _2+36072036 \sigma _0
   \sigma _1-128186058 \sigma _0 \sigma _2-275765490 \sigma _1 \sigma _2-10863916
   \sigma _0 \sigma _1 \sigma _2\nn
&\hspace{5mm}+N(111780240+4172771 \sigma _0+13985609
   \sigma _1+20407576 \sigma _2-36072036 \sigma _0 \sigma _1\nn
&\hspace{14mm}+128186058 \sigma _0
   \sigma _2+275765490 \sigma _1 \sigma _2+10863916 \sigma _0 \sigma _1 \sigma
   _2)\bigg]\e+O(\e^{3/2})
\,,\nn
\l_{\pa^4\f^2}&=\frac{7 \sigma _0}{3}+\frac{7 \sigma _1}{5}+\frac{7 \sigma _2}{15}+\frac{1}{421200 (6N-17)}\bigg[18292560+13346498 \sigma _0+10794218 \sigma _1\nn
   &\hspace{5mm}+1318945 \sigma _2+45236763 \sigma _0
   \sigma _1-13391664 \sigma _0 \sigma _2-39021840 \sigma _1 \sigma _2-1045856 \sigma
   _0 \sigma _1 \sigma _2\nn
   &\hspace{5mm}+N (-18292560-13655378 \sigma _0-10979546 \sigma
   _1-1380721 \sigma _2-45236763 \sigma _0 \sigma _1\nn
   &\hspace{14mm}+13391664 \sigma _0 \sigma
   _2+39021840 \sigma _1 \sigma _2+1045856 \sigma _0 \sigma _1 \sigma _2)\bigg]\e+O(\e^{3/2})\,,\nn
\l_{\f^3}&=\frac{N-1}{6N-17}\bigg(\frac{1259}{108}+\frac{105 \sigma _0}{143}+\frac{95 \sigma _1}{143}+\frac{1127 \sigma
   _2}{3861}+\frac{39 \sigma _0 \sigma _1}{4}+\frac{49 \sigma _0 \sigma
   _2}{12}\nn
   &\hspace{8mm}+\frac{21 \sigma _1 \sigma _2}{4}+\frac{7\sigma _0 \sigma _1 \sigma _2}{26} \bigg)\e+O(\e^{3/2})\,,\nn
\l_{\pa^2\f^3}&=-\frac{N-1}{6N-17}\bigg(\frac{3683}{1440}+\frac{301 \sigma _0}{6864}+\frac{469 \sigma _1}{10296}+\frac{931 \sigma _2}{34320}+\frac{947
   \sigma _0 \sigma _1}{480}+\frac{343 \sigma _0 \sigma
   _2}{360}\nn
   &\hspace{8mm}+\frac{91 \sigma _1 \sigma _2}{60}+\frac{119 \sigma _0 \sigma _1 \sigma
   _2}{4680}\bigg)\e+O(\e^{3/2})\,,
   \nn
\l_{\pa^4\f^3}&=\frac{N-1}{6N-17}\bigg(\frac{683333}{617760}+\frac{19733 \sigma _0}{14723280}+\frac{31703 \sigma
   _1}{22084920}+\frac{17983 \sigma _2}{14723280}+\frac{169757 \sigma _0 \sigma
   _1}{205920}+\frac{31997 \sigma _0 \sigma _2}{77220}\nn
   &\hspace{8mm}+\frac{35357 \sigma _1 \sigma
   _2}{51480}+\frac{2261 \sigma _0 \sigma _1 \sigma _2}{2007720}\bigg)\e+O(\e^{3/2})
\,.
\ee
For $\D\geq 24$, we obtain
\be
\l_\D&=-\frac{\sqrt{\pi } (\Delta -8) \Gamma \left(\frac{\Delta
   }{2}\right)}{2^{\Delta }(\Delta -16) (\Delta -14) (\Delta -12) \Gamma \left(\frac{\Delta
   -9}{2}\right)}\times\frac{(-1)^{\Delta/2}}{5054400}\times\frac{N-1}{6N-17}\nn
  &\times\Bigg[\frac{1}{6} (4054302720-2780559360 \Delta +696327936 \Delta ^2-96690816 \Delta
   ^3+8488048 \Delta ^4\nn
   &\quad-501120 \Delta ^5+19864 \Delta ^6-504 \Delta ^7+7 \Delta
   ^8)+\frac{1}{2} (1121402880-768890880 \Delta +171134208 \Delta ^2\nn
   &\quad-21080448
   \Delta ^3+1636624 \Delta ^4-86400 \Delta ^5+3112 \Delta ^6-72 \Delta ^7+\Delta
   ^8) \sigma _0 \sigma _1+\frac{1}{3} (704471040\nn
   &\quad-533191680 \Delta +132223488
   \Delta ^2-17512128 \Delta ^3+1440304 \Delta ^4-79920 \Delta ^5+2992 \Delta ^6-72
   \Delta ^7\nn
   &\quad+\Delta ^8) \sigma _0 \sigma _2+(301916160-256711680 \Delta
   +74562048 \Delta ^2-11412288 \Delta ^3+1076464 \Delta ^4\nn
   &\quad-66960 \Delta ^5+2752 \Delta
   ^6-72 \Delta ^7+\Delta ^8) \sigma _1 \sigma _2\Bigg]\e+O(\e^{3/2})
\,.
\ee
The anomalous dimensions of the bulk operators agree with the previous calculation \cite{Guo:2024bll}.
For $\D\geq 24$, the $N\ra\infty$ limit of $\l_{\D}$ in the Potts model corresponds to the $(-1)^{\D/2}$ part of \eqref{YL lambda k=3 result}.
The one-point function reads
\be\label{a phi3 k=3}
a_{\f^3}=\l_{\f^3}/C_{\f^3}=\;&\pm\frac{10}{(N+1)}\sqrt{-\frac{2}{231(6N-17)}}\bigg(\frac{1259}{108}+\frac{105 \sigma _0}{143}+\frac{95 \sigma _1}{143}+\frac{1127 \sigma
   _2}{3861}+\frac{39 \sigma _0 \sigma _1}{4}\nn
   &\hspace{18mm}+\frac{49 \sigma _0 \sigma
   _2}{12}+\frac{21 \sigma _1 \sigma _2}{4}+\frac{7\sigma _0 \sigma _1 \sigma _2}{26} \bigg)\e^{1/2}+O(\e)
\,.
\ee

\section{The bulk OPE coefficient of $\f^3$ in the Potts model}

Consider the matching condition
\be
\lim_{\e\ra 0}\a^{-1}\<\Box^k\f_a(x)\f_b(y)\f^3(z)\>
=\<K_a(x)\f_b(y)\f^3(z)\>_{\text f}
\,,
\ee
where we have defined $\f^3\equiv\sum_{a,b,c}^{N}T_{abc}\f_a\f_b\f_c$.
The action of $\Box^k$ was determined in Appendix A of \cite{Guo:2024bll}.
We find the bulk OPE coefficient of $\f^3$
\be\label{C}
C_{\f^3}=(-1)^k\frac{3(N-1)(N+1)^2}{4^k (k)_k^2}\a+O(\e)
\,.
\ee

\providecommand{\href}[2]{#2}\begingroup\raggedright\endgroup

\end{document}